\documentstyle[12pt,epsfig]{article}
\newcommand{\be}{\begin{equation}}
\newcommand{\ee}{\end{equation}}
\newcommand{\bea}{\begin{eqnarray}}
\newcommand{\eea}{\end{eqnarray}}

\begin{document}
\title{Antibaryon production in hot and dense nuclear matter
     \thanks{Supported by GSI Darmstadt.} }
\author{ W. Cassing \\
    Institut f\"{u}r Theoretische Physik, Universit\"{a}t Giessen \\
    D-35392 Giessen, Germany}
\maketitle
\begin{abstract}
The production of antibaryons is calculated in a microscopic
transport approach employing multiple meson fusion reactions
according to detailed balance relations with respect to
baryon-antibaryon annihilation. It is found that the abundancies
of  antiprotons as observed from peripheral to central collisions
of $Pb + Pb$ at the SPS and $Au + Au$ at the AGS can approximately
be described on the basis of multiple interactions of 'formed'
hadronic states which drive the system to chemical equilibrium by
flavor exchange or quark rearrangement reactions.
\end{abstract}

\vspace{2cm} \noindent PACS: 24.10.-i; 24.10.Cn; 24.10.Jv;
25.75.-q; 14.65.-q


\newpage

\section{Introduction}
Ever since the first observation of antiproton production in
proton-nucleus \cite{chamberlain,elioff,dorfan} and
nucleus-nucleus collisions \cite{JINR,BEVALAC1,BEVALAC2,KEK,GSI}
the production mechanism has been quite a matter of debate. Especially
in nucleus-nucleus collisions at subthreshold energies
traditional cascade calculations, that employ free $NN$ production
and $\bar{p}N$ annihilation cross sections, essentially fail in
describing the high cross sections seen from 1.5 - 2.1 A$\cdot$GeV
\cite{Batko91,Cass92,Faess1}. Thus multiparticle nucleon
interactions \cite{Danielewicz90,Weise} have been suggested as a
possible solution to this problem. On the other hand it has been
pointed out  that the quasi-particle properties of the nucleons
and antinucleons  might be important for the $\bar{p}$ production
process which become more significant with increasing nuclear
density. Schaffner et al. \cite{Schaffner91} found in a static
thermal relativistic model based on scalar and vector self
energies-- assuming kinetic and chemical equilibrium -- that the
$\bar{p}$-abundancy might be dramatically enhanced when assuming
the antiproton self energy in the medium to be given by charge conjugation of
the nucleon self energy. This assumption implies strong attractive
vector self energies for the antiprotons which lead to a
reduction of the necessary energy to produce $p\bar{p}$ pairs in the medium by
binary quasi-particle interactions. On the other hand, such self
energies will lead to different spectral slopes of protons and
antiprotons as pointed out in Ref. \cite{Koch}.

First nonequilibrium calculations within a fully relativistic
transport model for antiproton production -- including $\bar{p}$
annihilation as well as the change of the quasi-particle
properties in the medium -- have been performed in Ref.
\cite{Cass92}. There it was found that according to the reduced
antinucleon energy in the medium the threshold for
$\bar{p}$-production is shifted to lower energy and the antiproton
cross section prior to annihilation becomes enhanced e.g. for
$Si+Si$ at 2.1 GeV$\cdot$GeV by approximately a factor 70 as
compared to a relativistic cascade calculation where no in-medium
effects are incorporated. Later on, a couple of relativistic
transport calculations have been performed
\cite{Teis94,Ko93,RQMD,sibirtsev} essentially pointing out that
all the low energy data from proton-nucleus and nucleus-nucleus
collisions are compatible with attractive $\bar{p}$ self energies
(at normal nuclear matter density $\rho_0$) in the order of -100
to -150 MeV \cite{sibirtsev,Cass99,Ko96}. However, it had been
stressed at that time that the high antiproton yield might also be
attributed to mesonic production channels  \cite{ko1,ko2,Wittmann}
since $p \bar{p}$ annihilation leads to multi-pion final states
with an average abundancy of 5 pions \cite{Dover}, which e.g.
might stem from an intermediate state of 2 $\rho$-mesons and a pion.

With new data coming up on antibaryon production from
nucleus-nucleus collisions at the AGS \cite{AGSall,E877,AGSnew}
and SPS \cite{NA49,Na49b,NAxx,NA57,Andersen} the $\bar{p},
\bar{\Lambda}$ enhancement factors seen experimentally were no
longer that dramatic as at SIS energies, however, traditional
cascade calculations employing free production and annihilation
cross sections again were not able to reproduce the measured
abundancies and spectra \cite{AGS1,AGS2,Kahana,Bleicher,WA97c}
especially for $\Xi, \bar{\Xi}$ and $\Omega, \bar{\Omega}$. Here
additional collective mechanisms in the entrance channel have been
suggested such as color rope formation \cite{Sorge} or hot plasma
droplet formation \cite{Werner}. In another language this has been
addressed also as string fusion \cite{Carlos,Carlos2}, a precursor
phenomenon for the formation of a quark-gluon plasma (QGP).

The intimate connection of antibaryon abundancies with the
possible observation of a new state of the strongly interacting
hadronic matter, i.e. the quark-gluon plasma, has been often
discussed since the early suggestion in Ref. \cite{Rafelski} that
especially the enhanced yield of strange antibaryons --
approximately in chemical equilibrium with the other hadronic
states -- should be a reliable indicator for a new state of
matter. In fact, the data on strange baryon and antibaryon
production from the NA49 and WA97 Collaborations show an
approximate chemical equilibrium \cite{BM,becca,Redlich00,Red01}
with an enhancement of the $\Omega^-, \Omega^+$ yield in central
$Pb + Pb$ collisions (per participant) relative to $p Be$
collisions at the same invariant energy per nucleon by a factor
$\sim $ 15. As pointed out in Ref. \cite{Redlich2} the data on
multi-strange antibaryons at the SPS seem compatible with a
canonical ensemble in chemical equilibrium. At AGS energies of
11.6 A$\cdot$GeV/c, furthermore, a high ratio of
$\bar{\Lambda}$/$\bar{p}$ of $3.6^{+4.7}_{-1.8}$ has been reported
\cite{AGSnew} for central collisions of $Au + Au$, that is not
described by any approach so far. Strange flavor exchange
reactions \cite{Capella} help in creating multistrange
antibaryons, however, the latter strangeness enhancement factors
could not be described within traditional transport or cascade
simulations without additional assumptions such as enhanced string
tensions or reduced quark masses \cite{Bleicher}.

In a more recent paper Rapp and Shuryak have taken up again the
idea of multi-meson production channels for baryon-antibaryon
pairs \cite{Rapp} to describe the antiproton abundancies in
central $Pb+Pb$ collisions at the SPS by introducing additionally
a finite pion chemical potential which helps in enhancing the
multi-pion collision rate. Later on, Greiner and Leupold
\cite{Carsten} have applied the same concept for the
$\bar{\Lambda}$ production by a couple of mesons including a $K^+$
or $K^0$ (for the $\bar{s}$ quark). However, such estimates remain
schematic unless fully microscopic multi-particle calculations
support or disprove such suggestions. The problem here is that
most of the transport models include only binary reactions in the
entrance channel whereas the final channel of an energetic
collision may well consist of many hadrons emerging from the decay
of strings that are excited in the initial reaction. Thus, as has
been pointed out quite often \cite{Rapp,Brat,Bravina}, detailed
balance is not included on the many-particle level leading to an
improper equilibrium state for large times ($t \rightarrow
\infty$).

In this work we will address two separate questions: the first one
is of more formal nature and related to a transport approach that
properly takes into account reactions of 2 hadrons $\rightarrow n$
hadrons and vice versa employing detailed balance (Section 2). The
second one addresses a suitable covariant scheme for the
calculation of such multi-particle reactions in transport models.
The method and its implementation in the hadron-string-dynamics
(HSD) transport approach \cite{Cass99,Ehehalt} will be described
in Section 3. A first application of this novel approach is
devoted to the problem of antibaryon production in relativistic
nucleus-nucleus collisions by multi-meson reaction channels.
Respective calculations and studies at SPS and AGS energies --
with a focus on antiproton abundancies -- will be presented in
Section 4 whereas Section 5 concludes this work with a summary.

\section{Generalized transport equations}
In this Section a brief description of the relativistic transport
model is given with emphasis on a new development, i.e. the
multi-particle reaction dynamics. First we summarize (or review) the relevant
equations determining the dynamics of baryons and mesons and then
discuss a flavor rearrangement model for baryon-antibaryon
annihilation and production in the extended
HSD transport approach.

\subsection{Hadron transport and multi-particle transitions}

Since the covariant transport approach for binary reactions has
been extensively discussed in Refs. \cite{KLW1} and in the reviews
\cite{Cass99,Mal,Cassing90c} we only recall the basic equations
that are relevant for a proper understanding of the results to be
reported in this study.

For the discussion of the general collision terms the hadron self energies
will be discarded for transparency\footnote{The actual transport calculations,
however, include hadron self energies which optionally can be switched-off.},
such that the transport equation in the cascade limit reduces to
\begin{equation}
\label{vlasov}
p_\mu \partial^\mu_x  F_i(x,p) = ( p_0 \partial_t +
{\vec p} \cdot {\vec \partial}_r) F_i(x,p)  = I_{coll}^i ,
\end{equation}
where $F_i(x,p)$ is the Lorentz covariant 8 dimensional
phase-space distribution function for an off-shell hadron with quantum
numbers $i$, i.e.
\begin{equation}
\label{spectral}
F(x,p) = A(x,p) f(x,p),
\end{equation}
where $A(x,p)$ denotes the hadron spectral function while $f(x,p)$
describes the occupation probability in phase-space.
The off-shell propagation of hadrons leads to additional terms
in the l.h.s. of (1) that describe the change of the spectral
function $A_i(x,p)$ during the propagation. These terms are omitted here
since the actual calculations will be performed in the on-shell limit.
For details on the off-shell propagation of hadrons in
the medium the reader is
referred to Refs. \cite{Casju,Leupold}.

We now turn to a discussion of the collision term $I_{coll}^i$,
which includes the new elements to be presented below. In the most
general case it is a sum of collision integrals involving $n
\leftrightarrow m$ reactions,
\begin{equation}
\label{icoll}
I_{coll} = \sum_n \sum_m \ I_{coll} [n \leftrightarrow m].
\end{equation}
The general form for off-shell fermions with spectral functions
$A_i(x,p_i)$ in case of 2-body interactions is given by
\cite{Casju}

\bea \lefteqn{ I_{coll}^i [2 \leftrightarrow 2]  = } \nonumber \\
&& \frac{1}{2} \sum_j \sum_{k,l} \frac{1}{(2 \pi)^{12}} \int d^{4} p_{2} \,
d^{4} p_{3} \,
              d^{4} p_4 \ A_i(x,p) A_j(x,p_2) A_k(x,p_3)
              A_l(x,p_4)
\nonumber \\
  &&   \times W_{2,2}(p, p_2; i,j \mid p_3,p_4; k, l) \
 (2\pi)^4   \ \delta^{4}(p^{\mu} + p^{\mu}_2 - p_3^{\mu} - p_4^{\mu} )
\nonumber \\
    &&   \times   [ f_k(x,p_3) f_l(x,p_4)(1 - f_i(x,p))(1 -
    f_j(x,p_2))
\nonumber \\
     &&     -  f_i(x,p) f_j(x,p_2)(1 - f_k(x,p_3))(1-f_l(x,p_4)) ].
     \label{icoll2} \eea
This collision integral describes the change in the 8 dimensional
phase-space distribution function $F_i(x,p)$ due to the collisions
of two baryons with  momenta $p^{\mu}$ and $p^{\mu}_2$ and
discrete quantum numbers $i$ and $j$, respectively, whereas the
two fermions in the final state of the reaction  are labeled by
their momenta $p_3$ and $p_4$ and discrete quantum numbers $k$ and
$l$. The $\delta-$function guarantees energy and momentum
conservation in the individual collision while $ W_{2,2}(p, p_2;
i,j \mid p_3, p_4; k,l)$ denotes the transition probability (or
transition amplitude squared) for this reaction which in case of
fermions includes the antisymmetrization.

The generalization of (\ref{icoll2}) to $n \leftrightarrow m$ interactions is
straight forward and reads: \bea \lefteqn{I_{coll}^i [n
\leftrightarrow m]  = } \nonumber \\ && \frac{1}{2} N_n^m \sum_\nu
\sum_{\lambda} \left( \frac{1}{(2 \pi)^{4}} \right)^{n+m-1} \int
\left ( \prod_{j=2}^n    d^{4} p_j \,\ A_j(x,p_j)\right ) \left (
\prod_{k=1}^m d^{4} p_{k} \,
                A_k(x,p_k) \right ) \
\nonumber \\
  &&  \times  A_i(x,p) \  W_{n,m}(p, p_j; i,\nu \mid p_k; \lambda) \
 (2\pi)^4  \ \delta^{4}(p^{\mu} + \sum_{j=2}^n p^{\mu}_j - \sum_{k=1}^m p_k^{\mu})
\nonumber \\
   &&   \times    [\tilde{f}_i(x,p) \prod_{k=1}^m f_k(x,p_k) \prod_{j=2}^n \tilde{f}_j(x,p_j)
      -  f_i(x,p) \prod_{j=2}^n f_j(x,p_j) \prod_{k=1}^m \tilde{f}_k(x,p_k)].
\label{icollm}
\eea
In Eq. (\ref{icollm}) the quantities $\tilde{f}$ denote Pauli-blocking
or Bose-enhancement factors as
\begin{equation}
\tilde{f} = 1 + \eta f
\end{equation}
with $\eta$ = 1 for bosons and $\eta = -1$ for fermions,
respectively. The indices $\nu$ and $\lambda$
stand for the set of discrete quantum numbers in the initial
(except for particle $i$) and final states, respectively, and $N_n^m$
denotes a statistical factor that takes into account the number of
identical fermions and bosons in the initial and final states. In
the following we will assume that the transition probabilities
$W_{n,m}$ are evaluated with respect to asymptotically free
antisymmetrized fermion many-body
states and symmetrized meson states such that $N_n^m$ = 1.

In the on-shell quasi-particle limit the spectral functions,
furthermore, are given by\footnote{In general the spectral
functions for fermions differ from that of bosons \cite{Casju};
here  baryons and antibaryons with different spins are treated as
independent Klein-Gordon particles.}
\be
\label{spec} A_i(x,p) = 2 \pi \ \delta(p^2 - M_i^2) \ee which
leads to the 2-body collision integral for particle $i$ (in case of fermions) as:
\bea
\lefteqn{ I_{coll}^i [2 \leftrightarrow 2]  =  \frac{1}{2} \sum_j \sum_{k,l}
\frac{1}{(2 \pi)^{9}} \int \frac{d^{3} p_{2}}{2 E_2} \,
\frac{d^{3} p_{3}}{2 E_3} \, \frac{d^{3} p_4}{2 E_4} } \nonumber
\\
  &&   \times W_{2,2}(p, p_2; i,j \mid p_3,p_4; k, l) \
 (2\pi)^4   \ \delta^{4}(p^{\mu} + p^{\mu}_2 - p_3^{\mu} - p_4^{\mu} )
\nonumber \\
    &&    \times   [ f_k(x,p_3) f_l(x,p_4)(1 - f_i(x,p))(1 - f_j(x,p_2))
\nonumber \\
     &&     -  f_i(x,p) f_j(x,p_2)(1 - f_k(x,p_3))(1-f_l(x,p_4))],
\label{icoll2a} \eea where the energy $p^0_k$ is rewritten as
$E_k$. Note that the $p_0=E_1$ integration over the spectral
function $A_i(x,p)$ appears on both sides of the transport
equation and cancels out in the limit $\Gamma_i \rightarrow$ 0.
The  on-shell version of the collision integral
(\ref{icollm}) then reads

\bea \lefteqn{I_{coll}^i [n \leftrightarrow m]  =  \frac{1}{2} \sum_\nu
\sum_{\lambda} \left( \frac{1}{(2 \pi)^{3}} \right)^{n+m-1} \int
\prod_{j=2}^n    \frac{d^{3} p_j}{2 E_j} \, \prod_{k=1}^m
\frac{d^{3} p_{k}}{2E_k} \, } \nonumber \\
  &&   \times W_{n,m}(p, p_j;i,\nu \mid p_k; \lambda) \
 (2\pi)^4  \ \delta^{4}(p^{\mu} + \sum_{j=2}^n p^{\mu}_j - \sum_{k=1}^m p_k^{\mu})
\nonumber \\
   &&       [\tilde{f}_i(x,p) \prod_{k=1}^m f_k(x,p_k) \prod_{j=2}^n \tilde{f}_j(x,p_j)
      -  f_i(x,p) \prod_{j=2}^n f_j(x,p_j) \prod_{k=1}^m \tilde{f}_k(x,p_k)].
\label{icollma}
\eea
For large times $(t \rightarrow \infty$) all collision integrals
vanish, which implies that 'gain' and 'loss' terms become equal in
magnitude.

The number of reactions in the covariant  4-volume $d^3r dt$ = $dV
dt$ is obtained by dividing the gain and loss terms in the
collision integrals by the energy $p_0=E_1$, integrating over
$d^3p/(2\pi)^3$ and summing over the discrete quantum numbers $i$.
For the case of fermion two-body collisions this gives (using $p=p_1$) for
the 'loss' term

\bea \lefteqn{ \frac{d N_{coll} [2 \rightarrow 2]}{dt dV}  =
\sum_{i,j} \sum_{k,l} \frac{1}{(2 \pi)^{12}}  \int \frac{d^{3}
p_{1}}{2 E_1} \
 \frac{d^{3} p_{2}}{2 E_2} \,
\frac{d^{3} p_{3}}{2 E_3} \, \frac{d^{3} p_4}{2 E_4} }
\nonumber \\
  &&   \times  W_{2,2}(p_1, p_2; i,j \mid p_3,p_4; k, l) \
 (2\pi)^4   \ \delta^{4}(p^{\mu}_1 + p^{\mu}_2 - p_3^{\mu} - p_4^{\mu} )
\nonumber \\
  &&  \times  [   f_i(x,p_1) f_j(x,p_2)(1 - f_k(x,p_3))(1-f_l(x,p_4))].
\label{icoll2c} \eea
In case of $n \rightarrow m$ processes this leads to
\bea
\lefteqn{\frac{d N_{coll} [n \rightarrow m]}{dt dV}  =
\sum_{i,\nu} \sum_{\lambda} \left( \frac{1}{(2 \pi)^{3}}
\right)^{n+m}  \int \prod_{j=1}^n  \frac{d^{3} p_j}{2 E_j} \,
\prod_{k=1}^m   \frac{d^{3} p_{k}}{2E_k} \, } \nonumber \\
  &&   \times W_{n,m}( p_j; i,\nu \mid p_k; \lambda) \
 (2\pi)^4  \ \delta^{4}( \sum_{j=1}^n p^{\mu}_j - \sum_{k=1}^m p_k^{\mu})
       ( \prod_{j=1}^n f_j(x,p_j) \prod_{k=1}^m \tilde{f}_k(x,p_k))
\label{icollmc} \eea
and in case of $m \rightarrow n$ processes to
\bea \lefteqn{\frac{d N_{coll} [m \rightarrow n]}{dt dV}  =
 \sum_{i,\nu} \sum_{\lambda} \left( \frac{1}{(2
\pi)^{3}} \right)^{n+m}  \int \prod_{j=1}^n \frac{d^{3} p_j}{2
E_j} \, \prod_{k=1}^m \frac{d^{3} p_{k}}{2E_k} \, } \nonumber
\\
  &&  \times  W_{n,m}( p_j; i,\nu \mid p_k; \lambda) \
 (2\pi)^4  \ \delta^{4}( \sum_{j=1}^n p^{\mu}_j - \sum_{k=1}^m p_k^{\mu})
     ( \prod_{k=1}^m f_k(x,p_k) \prod_{j=1}^n
   \tilde{f}_j(x,p_j)) .
\label{icollmd} \eea For the phase-space configurations of
interest in this study the Pauli-blocking or Bose-enhancement
terms $\tilde{f_k}$ are $\approx 1$, which implies to replace the
quantum statistical ensembles by classical ones. In this limit
 the integrals over the final momenta can be carried out
provided that the transition probabilities $W_{n,m}$ do not
sensitively depend on the final momenta $p_k$. Employing the
definition of the $n-body$ phase-space integrals for total
4-momentum $P^\mu$ \cite{Byckling},
\be
\label{RN} R_n(P^\mu;M_1,..,M_n) = \left( \frac{1}{(2 \pi)^{3}}
\right)^{n} \int \prod_{k=1}^n \ \frac{d^{3} p_{k}}{2E_k} (2\pi)^4
\ \delta^{4}(P^{\mu} - \sum_{j=1}^n p^{\mu}_j ) , \ee one obtains
the recursion relation \cite{Byckling}
\be
R_n(P^\mu,M_1,..,M_n) =  \frac{1}{(2 \pi)^{3}}  \int
 \frac{d^{3} p_{n}}{2E_n} R_{n-1}(P^{\mu}- p^\mu_n; M_1,..,M_{n-1}). \ee
Note, that the phase-space integrals are of dimension
 GeV$^{2n-4}$ or (1/fm)$^{2n-4}$. Inserting (\ref{RN})
  this gives in case of $n \rightarrow m$
processes \bea \lefteqn{\frac{d N_{coll} [n \rightarrow m]}{dt dV}
= \sum_{i,\nu} \sum_{\lambda} \left( \frac{1}{(2 \pi)^{3}}
\right)^{n}  \int \left ( \prod_{j=1}^n   \frac{d^{3} p_j}{2 E_j}
\right ) \, } \nonumber \\
  &&   \times  W_{n,m}(P) \
 R_m(P^{\mu}=\sum_{j=1}^n p^{\mu}_j;M_1,..,M_m)
        \prod_{j=1}^n f_j(x,p_j) \nonumber \\
&& = \sum_{i,\nu} \sum_{\lambda} \left( \frac{1}{(2 \pi)^{3}}
\right)^{n}  \int \left ( \prod_{j=1}^n  d^{3} p_j \right ) \,
    P(n \rightarrow m)_{i,\nu}^\lambda  \
     \left (   \prod_{j=1}^n f_j(x,p_j)  \right )     ,
\label{icollm4} \eea where $M_1,..,M_m$ stand for the masses in
the final state, and in case of $m \rightarrow n$ processes \bea
\lefteqn{\frac{d N_{coll} [m \rightarrow n]}{dt dV}  =
 \sum_{i,\nu} \sum_{\lambda} \left( \frac{1}{(2
\pi)^{3}} \right)^{m}  \int \left ( \prod_{k=1}^m \frac{d^{3}
p_{k}}{2E_k} \right ) \, } \nonumber \\
  &&  \times  W_{n,m}(P) \
 R_n(P^{\mu} =\sum_{k=1}^m p_k^{\mu};M_1,...,M_n)
     \left ( \prod_{k=1}^m f_k(x,p_k) \right ) \nonumber \\
      &&  =
 \sum_{i,\nu} \sum_{\lambda} \left( \frac{1}{(2
\pi)^{3}} \right)^{m}  \int \left (\prod_{k=1}^m d^{3}
p_{k} \right ) \ P(m \rightarrow n)_\lambda^{i,\nu} \
     \left ( \prod_{k=1}^m f_k(x,p_k) \right ) .
\label{icollm5} \eea For fixed sets of quantum numbers $(i,\nu)$
and $\lambda$ in the initial and final states this relates the
integrands $P(n \rightarrow m)$ in (\ref{icollm4}) and $P( m
\rightarrow n)$ in (\ref{icollm5}) for individual scatterings as
(dropping the indices for quantum numbers):
\be
\label{Ratio} \frac{P(m \rightarrow n)}{P (n \rightarrow m)} =
\left[ \prod_{k=1}^m \frac{1}{2E_k}\right] \, \left[ \prod_{j=1}^n
{2 E_j}\right] \frac{R_n(P^{\mu} =\sum_{k=1}^n
p_k^{\mu};M_1,...,M_n)}{ R_m(P^{\mu}=\sum_{j=1}^m
p^{\mu}_j;M_1,..,M_m)}, \ee if $W_{n,m}$ essentially depends only
on the invariant energy $\sqrt{s} = \sqrt{P^2}$. Note, that the
r.h.s. of (\ref{Ratio}) is in units of GeV$^{3(n-m)}$ or
fm$^{3(m-n)}$ such that a factor $(dV)^{n-m}$ is needed to
interpret the quantities as relative 'probabilities'.

Thus, once the transition probabilities $W_{n,m}$ are known as a
function of $\sqrt{s}$ for a given set of quantum numbers, the
integrand $P(n \rightarrow m)$ in (\ref{icollm4}) is determined by
phase space and the backward reactions in (\ref{icollm5}) are
fixed by Eq. (\ref{Ratio}).

\subsection{Antibaryon annihilation and recreation}
In the following the processes $B \bar{B} \leftrightarrow m \
mesons$ are discussed, which are of relevance for annihilation of
antibaryons on baryons and the recreation of $B\bar{B}$ pairs by
$m \ meson$ interactions. The 4-differential collision rate for
baryon-antibaryon annihilation $(1+2 \rightarrow 3, ..,m+2)$ then
is given by \bea \lefteqn{\frac{d N_{coll} [B\bar{B} \rightarrow m
\ mesons]}{dt dV}  = \sum_{i,j} \sum_{\lambda_m} \left(
\frac{1}{(2 \pi)^{3}} \right)^{2}  \int   \frac{d^{3} p_1}{2 E_1}
\, \frac{d^{3} p_2}{2 E_2} \,} \nonumber \\
  &&  \times  W_{2,m}(P= p_1+p_2;i,j;\lambda_m) \
 R_m(P^\mu;M_3,..,M_{m+2})
        f_i(x,p_1) f_j(x,p_2).
\label{ic1}
\eea
The integrand is related to the annihilation cross section
$\sigma_{ann.}(\sqrt{s})$ for a baryon-antibaryon pair with quantum numbers $i,j$ as
\cite{Byckling}
$$\sum_m \sum_{\lambda_m} \ W_{2,m}(p_1 + p_2;i,j;\lambda_m) \
 R_m(p_1^\mu+ p_2^\mu;M_3,..,M_{m+2}) $$
\be
\label{cross}
 = 2 \sqrt{\tilde{\lambda}(s,M_1^2,M_2^2)}\ \sigma_{ann.}(\sqrt{s}) =
 4 E_1 E_2 \  v_{rel} \ \sigma_{ann.}(\sqrt{s})
 \ee
with the Lorentz-invariant relative velocity
\cite{Byckling,Lang93}
\be
\label{vrel} v_{rel}= \frac{\sqrt{\tilde{\lambda}
(s,M_1^2,M_2^2)}}{2 E_1 E_2}, \ee involving
\be
\tilde{\lambda}(x,y,z) = (x-y-z)^2 - 4 yz.
\ee
In (\ref{cross}) the sum runs over the final meson multiplicity
($m$ $\approx$ 2, ... , 9) in the
final state and over $\lambda_m$ which denotes all discrete quantum
numbers of the final mesons for given multiplicity $m$.

Note, that by summing (\ref{ic1}) additionally over $m$, but
keeping the quantum numbers $i,j$ fixed, one arrives at
\bea
\frac{d N_{coll}^{i,j}}{dt dV}  =
 \frac{1}{(2 \pi)^{6}}
 \int   d^{3} p_1\,  d^{3} p_2 \ v_{rel}(p_1,p_2) \
 \sigma_{ann}(\sqrt{s}) \
       f_i(x,p_1) f_j(x,p_2).
\label{ic1b} \eea If the product of the relative velocity and the
cross section, i.e. $v_{rel} \sigma_{ann}$, is approximately constant (see below) the
integrals over the momenta in (\ref{ic1b}) give the classical
Boltzmann limit \bea \frac{d N_{coll}^{i,j}}{dt dV}  = \ < v_{rel}
\ \sigma_{ann} > \ \rho_i(x) \rho_j(x), \label{ic1c} \eea where
$\rho_i(x)$ is the density of the hadron with quantum numbers $i$.

The number of reactions per volume and time for the back processes
is then given by ($\lambda_m= k_1,..,k_m)$
\bea \lefteqn{\frac{d
N_{coll} [m \ mesons \rightarrow B\bar{B}]}{dt dV}  = \sum_{i,j}
\sum_{\lambda_m} \left( \frac{1}{(2 \pi)^{3}} \right)^{m}  \int
\left (\prod_{k=3}^{m+2} \frac{d^{3} p_{k}}{2E_k} \right ) \, }
\nonumber
\\
  &&  \times  W_{2,m}(\sqrt{s};i,j,\lambda_m) \
 R_2(P^\mu=\sum_{k=3}^{m+2} p_k^\mu;M_1,M_2)
    \left (  \prod_{k=3}^{m+2} f_k(x,p_k) \right ) ,
\label{ic5}
\eea
assuming $W_{2,m}(\sqrt{s};i,j,\lambda_m)$ to depend only on the available
energy $\sqrt{s}$ and conserved quantum numbers.

To proceed further, some simplifying assumptions have to be
invoked to lead to a tractable problem for antibaryon annihilation
and production. Experimentally, the differential multiplicity in the pions from
$p \bar{p}$ annihilation at low $\sqrt{s}$ above threshold can be described as
\be
\label{pion5} P(N_\pi) \approx \frac{1}{\sqrt{2\pi} D} \exp(-
\frac{(N_\pi - <N_\pi >)^2}{2 D^2}) \ee with an average pion
multiplicity of $<N_\pi> \approx$ 5 and $D^2$ = 0.95 \cite{Dover}.

This observation is  reminiscent of flavor rearrangement processes
in the $B\bar{B}$  annihilation reaction to vector mesons and
pseudoscalar mesons, e.g. $\rho + \rho + \pi$ or $\omega + \omega
+ \pi^0$, where the $\rho$ and $\omega$ 'later' decay to 2 or 3
pions, respectively. In this picture the $\rho + \rho + \pi$ final
channel in $p\bar{p}$ annihilation is the dominant process leading
finally to 5 pions. Alternatively, the $\omega + \omega + \pi^0$
channel leads to 7 pions in the final channel which will appear on
the scale of the $\omega$-meson lifetime. Three pions are obtained
in the direct 3 pion decay which, however, is substantially
suppressed at higher $\sqrt{s}$ due to spin multiplicities (see
below).

For the problem of interest we thus employ a quark rearrangement
model for $B\bar{B}$ annihilation to 3 mesons as illustrated in
Fig. \ref{bild1}, where the final mesons $M_i$ may be pseudoscalar
or vector mesons, i.e ($\pi, K, \eta$) or $(\rho, \omega, K^*,
\phi)$, respectively. In this model there is no creation of an
$s\bar{s}$ pair for nucleon-antinucleon annihilation due to the
conservation of constituent quark flavors. As can be extracted
from the detailed experimental cross sections given in Ref.
\cite{LB}, such processes are suppressed by more than an order of
magnitude. In principle, such channels can additionally be
included in the model, however, the backward reactions then are
also suppressed by the same relative factor according to detailed
balance. For the study of interest such 'small' channels are
discarded.

In the following, the quantum numbers denoted by $\lambda_m$ will
be separated into different channels $c$, that can be
distinguished by their mass decomposition, and degenerate quantum
numbers such as spin multiplicities and isospin projections. In
the latter sense the sum over the final quantum numbers
$\lambda_m$ in (\ref{ic1}), (\ref{ic5}) then includes a sum over
the mass partitions $c=(M_3,M_4,M_5)$, a sum over the spins of
 the mesons and a sum over all isospin quantum numbers, that
are compatible with charge conservation in the transition. The
probability for a channel
$c= (M_3,M_4,M_5)$ then reads
\be \label{model}
P_c(\sqrt{s};M_3,M_4,M_5) = N_3(\sqrt{s})\  R_3(\sqrt{s};M_3,M_4,M_5)\
N_{fin}^c,
\ee
where the number of 'equivalent' final states in the channel $c$ is given by
\be
\label{model1} N_{fin}^c = (2s_3+1)(2s_4+1)(2s_5+1)
\frac{F_{iso}}{N_{id}!} . \ee
 In (\ref{model1})
$s_j$ denote the spins of the final mesons, $F_{iso}$ is the
number of isospin projections compatible with charge conservation
while $N_{id}$ is the number of identical mesons in the final
channel (e.g. $N_{id}$ = 3 for the $\pi \pi \pi$ final channel).
This combinatorial problem for the final number of states
$N_{fin}^c$ is of finite dimension and easily tractable
numerically. For each mass partition $c=(M_3,M_4,M_5)$ the decay
probability then is given by the 3-body phase space and the
allowed number of final states $N_{fin}^c$ since the absolute
normalization -- described by $N_3(\sqrt{s})$ -- is fixed by the
constraint $\sum_c P_c = 1$.

As an example let us consider the problem of nucleon-antinucleon
annihilation, where the following final meson channels contribute,
\be \label{index} (1) \ \pi \pi \pi \ (2) \ \pi \pi \rho \ (3) \
\pi \pi \omega \ (4)\ \pi \rho \rho \ (5)\ \pi \rho \omega \ (6) \
\pi \omega \omega  , \ee excluding 3 vector mesons in the final
channel. According to (\ref{model}) the distribution in the final
number of pions (including the explicit vector meson decays to
pions) can be evaluated as a function of $\sqrt{s}$ since it only
depends on the phase space and the number of possible final states
$N_{fin}^c$ in each channel $c$. The numerical results are
displayed in Fig. \ref{bild2}
 for 2.3 GeV $\leq
\sqrt{s} \leq$ 4  GeV in comparison to the parametrization
(\ref{pion5}) (solid line). Here the horizontal bars indicate the
range of $N_\pi$-pion probabilities
when varying $\sqrt{s}$ from 2.3 to 4 GeV. Obviously, the simple
phase-space model (\ref{model})
is in a fair agreement with the experimental observation. For
related or more extended models for $p\bar{p}$ annihilation the
reader is referred to Ref. \cite{Dover}.

For the backward reactions, i.e. the 3 meson fusion to a
$B\bar{B}$ pair, the quarks and antiquarks are redistributed in a
baryon and antibaryon, respectively, incorporating the baryons $N,
\Delta, \Lambda, \Sigma, \Sigma^*, \Xi, \Xi^*, \Omega$ as well as
their antiparticles. In line with (\ref{model}) the relative
population of states (with the same quark content) is determined
by phase space, i.e. $$ P_{c'}(\sqrt{s};M_1,M_2) = N_2(\sqrt{s}) \
R_2(\sqrt{s};c'=(M_1,M_2))\ (2s_1+1)(2s_2+1) = $$ \be \label{back}
N_2(\sqrt{s}) \ R_2(\sqrt{s};M_1,M_2) \ N_{B}^{c'}, \ee where
$N_B^{c'}$ now denotes the number of final states for the
particular mass channel $c'$ in the backward reaction. The
absolute normalization $N_2(\sqrt{s})$ is fixed again by the
constraint $\sum_{c'} P_{c'}$ = 1.

As an example consider the reactions $\pi^- \pi^+ \pi^-$ or $\pi^- \rho^+ \pi^-$
or $\pi^- \rho^+ \rho^-$ (and isospin combinations), i.e.
 $\bar{u}d + \bar{d}u + \bar{u}d \rightarrow (\bar{u} \bar{u}
\bar{d}) + (u d d)$: here the final states may be either $\bar{p}
+ n$, $\bar{\Delta}^- + n$, $\bar{p} + \Delta^0$ or $\bar{\Delta}^- + \Delta^0$
within the Fock space considered. Note,
that the final states with a $\Delta$-resonance  are favored due to the spin
factors in (\ref{back}), however, somewhat suppressed by the 2-body
phase-space integral $R_2(\sqrt{s})$ for low $\sqrt{s}$.

One is thus left with the $B\bar{B}$ annihilation problem
\bea \lefteqn{ \frac{d N_{coll}
[B\bar{B} \rightarrow 3 \  mesons]}{dt dV}  =  \sum_{c}
\sum_{c'}  \frac{1}{(2 \pi)^{6}}   \int
\frac{d^{3} p_1}{2 E_1} \,  \frac{d^{3} p_2}{2 E_2} \  W_{2,3}(\sqrt{s})} \nonumber
\\
  &&  \times N_3(\sqrt{s})  \
 R_3(p_1 + p_2;c=(M_3,M_4,M_5))\ N_{fin}^c
       \ f_i(x,p_1) f_j(x,p_2),
\label{ic6} \eea where $(M_1, M_2)$ denote the baryon and
antibaryon masses in the channel $c'$  and $(M_3,M_4,M_5)$ the
final meson masses in the channel $c$. Eq. (\ref{ic6}) can be
rewritten as \bea \lefteqn{\frac{d N_{coll} [B\bar{B} \rightarrow
3 \ mesons]}{dt dV}  = } \nonumber \\ && \sum_{c} \sum_{c'}
\frac{1}{(2 \pi)^{6}} \int d^3p_1 \ d^3p_2 \
P^{2,3}_{cc'}(\sqrt{s})    \ f_i(x,p_1) f_j(x,p_2) \label{ic6b}
\eea with the channel probabilities
\be
 P^{2,3}_{cc'}(\sqrt{s}) =  \frac{1}{4E_1 E_2} \,
    W^{2,3}(\sqrt{s}) \ N_3(\sqrt{s}) \
 R_3(p_1 + p_2;(M_3,M_4,M_5))\ N_{fin}^c .
 \label{p2}
 \ee
 Note, that by construction we have
 \be
 \sum_c  P^{2,3}_{cc'}(\sqrt{s}) = \frac{1}{4E_1 E_2} \,
    W^{2,3}(\sqrt{s}) = v_{rel} \
 \sigma_{ann}(\sqrt{s})_{c'},
 \label{p3}
 \ee
 where $v_{rel}$ denotes the relative velocity (\ref{vrel}) and
 $\sigma_{ann}(\sqrt{s})_{c'}$ is the total annihilation cross
 section for $B\bar{B}$ pairs of channel $c'$.

The backward invariant collision rate is given by
\bea \lefteqn{\frac{d N_{coll} [3 \ mesons \rightarrow B\bar{B}]}{dt dV}  =
 \sum_{c} \sum_{c'}  \frac{1}{(2
\pi)^{9}}   \int \left (\prod_{k=3}^5 \frac{d^{3} p_{k}}{2E_k}
\right ) \ W_{2,3}(\sqrt{s})} \nonumber \\
  &&   \times  N_2(\sqrt{s}) \
 R_2(\sum_{k=3}^5 p_k;c'=(M_1,M_2)) \ N_B^{c'}
     \left ( \prod_{k=3}^5 f_k(x,p_k) \right )  .
\label{ic7}
\eea
Using Eq. (\ref{Ratio}), the relation (\ref{cross}) for 3 mesons in the final state
and (\ref{p3}) one arrives at

\bea \lefteqn{\frac{d N_{coll} [3 \ mesons \rightarrow B\bar{B}]}{dt dV}  =
 \sum_{c} \sum_{c'} \frac{1}{(2
\pi)^{9}}   \int
\frac{d^{3} p_{3}}{2E_3} \, \frac{d^{3} p_{4}}{2E_4} \, \frac{d^{3} p_{5}}{2E_5} \,
 4 E_1 E_2 \  }
\nonumber \\
  &&  \times  v_{rel}\ \sigma(\sqrt{s})_{c'}  \
  \frac{N_2(\sqrt{s})}{N_3(\sqrt{s})}
 \frac{R_2(P^\mu;c'=(M_1,M_2))}{R_3(P^\mu ;c=(M_3,M_4,M_5))} \frac{N_B^{c'}}{N_{fin}^c} \nonumber \\
  && \hspace{3cm} \times  f_3(x,p_3)  f_4(x,p_4)  f_5(x,p_5)
 \label{final}
 \eea
 for the backward reaction $3+4+5 \rightarrow 1+2$.
Eq. (\ref{final}) can now be rewritten as \bea \lefteqn{\frac{d
N_{coll} [3 \ mesons \rightarrow B\bar{B}]}{dt dV}  = } \nonumber
\\ &&
  \sum_{c} \sum_{c'}  \frac{1}{(2
\pi)^{9}}   \int d^3 p_3 \ d^3p_4 \ d^3p_5 \
 P^{3,2}_{cc'}(\sqrt{s}) \
     f_3(x,p_3)  f_4(x,p_4)  f_5(x,p_5)
 \label{finalb}
 \eea
 with the 'transition integrand'
 \be
 P^{3,2}_{cc'}(\sqrt{s}) =
 \frac{ E_1 E_2}{2 E_3 E_4 E_5} v_{rel}\ \sigma(\sqrt{s})_{c'}  \   \frac{N_2(\sqrt{s})}{N_3(\sqrt{s})}
 \frac{R_2(P;c'=(M_1,M_2))}{R_3(P;c=(M_3,M_4,M_5))} \ \frac{N_B^{c'}}{N_{fin
 }^c},
 \label{finalc}
 \ee
which is of dimension GeV$^{-3}$ or fm$^3$.

\section{Numerical implementation}
For a reformulation of the 'transition integrands' (specified in (\ref{finalc})) in a
test-particle representation one has to recall
that the average density of a meson with quantum numbers $k$ is obtained
by integration over momentum as:
\be
n_k(x) = \frac{1}{(2 \pi)^{3}} \int d^3p \ f_k(x,p), \ee where
e.g. charge, strange flavor content, total spin and spin
projection are specified by the discrete quantum number $k$. The
conversion formula thus reads:
\be
\label{conv}
\frac{1}{(2 \pi)^{3}} \int d^3p \ f_k(x,p) \rightarrow
\frac{1}{dV} \sum_{i \ \epsilon \ dV},
\ee
where $dV$ is a (small) finite volume and the sum runs over all
test particles in the volume $dV$ with quantum numbers $k$. The
number of $B \bar{B}$ annihilations in the volume $dV$ during the
time $dt$ is thus given by \cite{Lang93}
\be
\label{twobody}
N_{B \bar{B}} = \frac{dt}{dV} \sum_{i,j \ \epsilon \ dV}
v_{rel}(i,j) \sigma_{ann}(\sqrt{s}_{i,j})
\ee
with the invariant energy squared
\be
s_{i,j} = (p_1 + p_2)^2, \ee where $p_1, p_2$ denote the 4-momenta
of the colliding $B\bar{B}$ pair. The relative velocity
$v_{rel}(i,j)$ is given by (\ref{vrel}) while the annihilation
cross section $\sigma_{ann}(\sqrt{s})$, furthermore, has to be
specified for all baryon-antibaryon pairs. This cross section is
rather well known for nucleon-antinucleon reactions
\cite{Dover,PDG,LB}, however, the channels involving $\Lambda,
\Sigma, \Sigma^*, \Xi, \Xi^*, \Omega^-$ baryons or their
antiparticles are not available experimentally by now and have to
be modeled to some extent.

For guidance we recall that the product $v_{rel} \
\sigma_{ann}(\sqrt{s}) \approx$ 50 mb for a wide range of
energies $\sqrt{s}$ in case of $p\bar{p}$ annihilation. This is
demonstrated explicitly in Fig. \ref{bild1b}, where the
experimental annihilation cross section for $p\bar{p}$ from Ref.
\cite{LB} is compared to the approximation
\begin{equation}
\label{appann} \sigma_{ann} (\sqrt{s}) = \frac{50 [mb]}{v_{rel}} ,
\end{equation}
which holds well in the dynamical range of interest.
We thus can adopt the Boltzmann limit (\ref{ic1c}) to estimate the
$B\bar{B}$ annihilation time at nucleon density $\rho$ as
\begin{equation}
\label{estimate2} \tau_{ann.} \approx (5 fm^2 \rho)^{-1} \approx
1.2 \ \frac{\rho_0}{\rho}\  [fm/c].
\end{equation}

In this work we will proceed with dynamical calculations in the
strangeness sector $S$=0, thus essentially addressing the
$\bar{p}$ abundancies in relativistic nucleus-nucleus collisions.
In this case the approach formulated above does not involve any
new parameter or cross section; it is just an extension of the HSD
approach \cite{Cass99,Ehehalt,Geiss} to include the $B\bar{B}
\leftrightarrow$ 3 meson reactions by detailed balance.

The number of backward reactions by 3 mesons in the test-particle
picture in the volume $dV$ and time $dt$ according to
(\ref{final}) for a given mass channel $c'$ is given by $$N_{3
meson} = \frac{dt}{dV dV} \sum_{i,j,k \ \epsilon \  dV} \frac{E_1
E_2}{2 E_i E_j E_k} v_{rel}(1,2) \sigma(\sqrt{s})_{c'} $$  \be
\label{trans}
\times \frac{N_2(\sqrt{s})}{N_3(\sqrt{s})}
 \frac{R_2(\sqrt{s};c'=(M_1,M_2))}{R_3(\sqrt{s};c=(M_i,M_j,M_k))}
\frac{N_B^{c'}}{N_{fin}^c} = \sum_{i,j,k \ \epsilon \ dV} \
P_{ijk}, \ee where the channel $c$ is defined by the colliding
mesons (cf. (\ref{ic5})) and the outgoing channel $c'$ by the
$B\bar{B}$ pair with masses $M_1$ and $M_2$ and energies $E_1$ and
$E_2$, respectively. In (\ref{trans}) the summation over the
mesons in the volume $dV$ is restricted to $i < j < k$ in case of
3 identical mesons (e.g. 3 pions) and to $i < j$ in case of 2
identical mesons $i,j$ in order to account for the statistical
factor $N_{id}!$ in Eq. (\ref{model1}).

 Eqs. (\ref{twobody}) and (\ref{trans}) are well suited for a Monte Carlo decision
problem, i.e. a transition is accepted if the probability
$P_{ijk}$ is larger than some random number in the interval [0,1].
One has to assure only, that all $P_{ijk}$ are smaller than 1,
which -- for a fixed volume $dV$ -- can easily be achieved by
adjusting the time-step $dt$. This evaluation of scattering
probabilities is Lorentz-invariant and does not suffer from
geometrical collision criteria as in the standard approaches
\cite{Wolf90,Batko3}, that imply a different sequence of
collisions when changing the reference frame by a Lorentz
transformation \cite{kodama}. For $2 \leftrightarrow 2$
transitions it has first been employed and tested by Lang et al.
in Ref. \cite{Lang93}; this method is also implemented in the HSD
approach, where it can be used optionally instead of the standard
geometrical collision criteria as described e.g. in Refs.
\cite{Cassing90,Wolf90,Bertsch88}. The present implementation in
this respect is a straight forward generalization of the concept
in Ref. \cite{Lang93} to $2 \leftrightarrow 3$ reactions.  It is
worth to point out that this numerical implementation is a
promising way to treat $n \leftrightarrow m$ transitions in
transport theories without violating covariance or causality. In
case of infinitesimal volumes $dV$ and time steps $dt$ it gives
the correct solution to the many-particle Boltzmann equation.

For the actual numerical calculations a dynamical time-step size
$dt$ is employed which on average amounts to $dt \approx
0.5/\gamma_{cm}$ [fm/c], where $\gamma_{cm}$ is the Lorentz factor
in the nucleus-nucleus cms, i.e. $\gamma_{cm} \approx$ 9.3 for
collisions of $Pb + Pb$ at 160 A$\cdot$GeV. The volume $dV$ is
chosen to be $dV = A \ dz$ with the transverse area $A = 9 fm^2$
and $dz = 3/\gamma_{cm}$ [fm]. Variations of these parameters
within a factor of 3 do not change the numerical results to be
presented below. In order to avoid numerical artefacts, that are
due to the finite volume $dV$, a $B\bar{B}$ pair, that has been
produced by meson fusion, is not allowed to annihilate on each
other again without performing an additional collision in between.
On the other hand, the mesons stemming from a particular
$B\bar{B}$ annihilation are not allowed to fuse again with the
same partners for the backward reaction, if no intermediate extra
collision has occured.

As a numerical test the number of collisions in a single box of
volume 10 fm$^3$ during the time $dt$ =1 fm/c has been calculated
with spatially uniform phase-space distributions given by a
classical system of hadrons in thermal and chemical equilibrium,
i.e.
\be
\label{equil} f_k(p) = \frac{(2s+1)(2I+1)}{(2\pi)^3}
{\exp(-E_k(p)/T)}  \ee with  $s$ and $I$ denoting spin and
isospin, respectively. The particles taken into account are $N,
\Delta$ and their antiparticles and $\pi, \rho, \omega$ on the
meson side in the strangeness sector $S$=0. The numerical results
for the number of $B\bar{B}$ annihilation collisions ($\rightarrow
\pi \rho \rho$) are shown in Fig. \ref{bild3} in terms of the
dashed line as a function of $\sqrt{s}$, which corresponds to the
invariant energy in an individual collision. As can be seen from
Fig. \ref{bild3} the dashed line very well coincides with the
solid line that corresponds to the energy differential number of
$\pi \rho \rho$ collisions for the backward reactions. Thus the
numerical scheme employed well reproduces the detailed balance
relation in thermal equilibrium for a given channel combination
$cc'$. Without explicit representation we mention that the
detailed balance relation is fulfilled for all channel
combinations $cc'$ specified above.

\section{Nucleus-nucleus collisions}
\subsection{SPS energies}
The most complete set of data on antibaryon production in
nucleus-nucleus collisions is available from the NA44 \cite{NA44},
NA49 \cite{Na49b} and WA97 \cite{NAxx,NA57} Collaborations for
$Pb+Pb$ collisions at SPS energies of 160 A$\cdot$GeV, that allow
for stringent tests of the dynamics proposed. Since the extension
of the HSD transport approach is described in the previous Section
and no new parameters enter into the calculations, we proceed with
the actual results.

\subsubsection{Cascade calculations}
Though antiproton self energies have been found to be important in
nucleus-nucleus collisions at subthreshold energies
\cite{sibirtsev,Cass99,Ko96}, we start with cascade calculations
because the initial invariant energy per nucleon is large compared
to the $B\bar{B}$ threshold. However, before coming to the
antibaryon abundancies the performance of the HSD transport
approach has to be tested in comparison to experimental data for
baryons and mesons. Since detailed differential
spectra for protons, hyperons, pions and kaons have been presented
in Ref. \cite{Geiss} in comparison to the experimental data for
central collisions of $Pb+Pb$ at 160 A$\cdot$GeV, we concentrate
here on particle abundancies as a function of 'centrality'. The
latter is defined by the number of participants $A_{part}$, which
is extracted from the transport calculation at impact parameter
$b$ as
\be
A_{part}(b) = 2 A - N_0(b), \ee
where $N_0(b)$ is the number of
nucleons that were not involved in any hard scattering process.

The average number of charged pions $<\pi> = (<\pi^+ + \pi^->)/2$
(divided by $A_{part}$) is shown in Fig. 5 as a function of $A_{part}$
in comparison to the data
from Ref. \cite{Na49b} for $Pb+Pb$ at 160 A$\cdot$GeV. The number
of pions per participant nucleon is found to be approximately
constant within 10 \% as a function of centrality; there is a
slight trend in the data as well as in the HSD calculations (solid
line) for an increase of $<\pi>/A_{part}$ for central collisions.
However, the HSD transport calculations overestimate the pion
abundancy by about 20 \%. Such an overprediction of pion
multiplicities in heavy systems occurs in other transport
approaches as well \cite{Bravina,Bass98} -- or is even higher --
and is not well understood so far. At SIS energies of 1--2
A$\cdot$GeV it has been argued in Ref. \cite{Larionov} that a
quenching of nucleon resonances at baryon densities $\rho \ge
\rho_0$ might be responsible for the relative pion suppression
seen experimentally in heavy systems, but it is not yet clear if
such a mechanism will also explain the discrepancies at SPS
energies. We thus have to keep in mind this overprediction of
pions especially when comparing particle ratios (see below).

The effect of antibaryon annihilation and reformation by means of
meson fusion channels is shown in Fig. 6, where the
$<\bar{p}>/<\pi>$ ratio is displayed for $Pb+Pb$ at 160
A$\cdot$GeV as a function of $A_{part}$. The dashed line shows a
calculation without including annihilation channels in the
transport calculation, the dotted line gives the results including
the annihilation to mesons while the solid line is obtained when
including both, annihilation and meson fusion channels by detailed
balance. The first point in Fig. 6 gives the numerical result for
$NN$ interactions at $T_{lab}$ = 160 GeV ($A_{part}$ = 2). As
shown in Table 1 of Ref. \cite{Geiss}, the description of the HSD
approach for $pp$ interactions is well in line with the data on
$\pi^+, \pi^0, \pi^-, K^+, K^-, K^0_s$, $\Lambda+\Sigma^0$,
$\bar{\Lambda} + \bar{\Sigma}^0$ as well as $p$ and $\bar{p}$
multiplicities at SPS energies, such that this point may serve for
reference to the particle abundancies in the elementary $NN$
interaction. As seen from Fig. 6 the $<\bar{p}>/<\pi>$ ratio
slightly drops with centrality for peripheral reactions, however,
stays approximately constant for $A_{part} \ge $ 150 when
neglecting annihilation. Thus antiprotons in central $Pb+Pb$
collisions are produced less frequent than pions by about 30 \%
relative to the $NN$ interaction in vacuum. The effect of
annihilation (dotted line) sets in already for peripheral
reactions and amounts to a factor $\sim$6--7 suppression for
central collisions. The latter suppression is sensitive to the
formation time $\tau_F$ of the antibaryons and becomes larger
(smaller) for shorter (longer) $\tau_F$. In the actual
calculations we have used $\tau_F$ = 0.8 fm/c for all hadrons
\cite{Geiss}. When including annihilation as well as meson fusion
channels by detailed balance, the $<\bar{p}>/<\pi>$ ratio (solid
line) becomes again close to the calculation that does not include
annihilation nor meson fusion reactions to $B\bar{B}$ (dashed
line). Thus on average the annihilation channels are almost
compensated by the reproduction channels.

In order to get some idea about the dynamical origin of the
approximately constant $\bar{p}/\pi$ ratio shown in Fig.
\ref{6new} by the solid line the reaction rate $B+\bar{B}
\rightarrow mesons$ (dashed histogram in Fig. \ref{bild5} ) is
compared to the backward reaction rate (solid histogram in Fig.
\ref{bild5}) for a central collision of $Pb+Pb$ at 160
A$\cdot$GeV. Fig. \ref{bild5} demonstrates that both rates are
comparable within the statistics. Thus an approximate local
chemical equilibrium is established very fast between the
nonstrange antibaryon degrees of freedom and the nonstrange mesons
$\pi, \rho$ and $\omega$. The latter fact is supported by the
absolute number of reactions $B\bar{B} \rightarrow mesons$ which
is about 4-5 times higher than the final number of antibaryons per
event. On the other hand, the number of backward reactions is
$\sim$ 96\% of the number of annihilation reactions leading to a
small net absorption of the antibaryons produced initially by
baryon-baryon ($\sim$ 73\%) or meson-baryon ($\sim$ 27\%)
inelastic collisions. Only $\sim 12$\% of the final antibaryons
stem from 'hard' baryon-baryon or meson-baryon reactions; the
dominant amount of final $\bar{p}$'s ($\sim$ 88\%) are from the 3
meson fusion reactions indicating that the memory with respect to
the initial 'hard' collision phase is practically lost.

One might worry about the sensitivity of these results to the
annihilation cross section $\sigma_{ann}(\sqrt{s})$ that so far
has been taken to be the free cross section (in the
parametrization from Ref. \cite{sibirtsev}). However, this
quantity might change in the medium due to screening effects. It
is clear that any enhancement of this cross section will lead to
an even faster equilibration in the light flavor degrees of
freedom and to a more perfect chemical equlibrium. Thus numerical
calculations have been performed for central $Pb+Pb$ collisions at
160 A$\cdot$GeV by assuming that $\sigma_{ann}$ is reduced by a
factor of 2. This leads to a reduction of the total number of
annihilation reactions (and backward reactions) by a factor of 2
-- which essentially reduces the numerical statistics -- but
leaves the conclusions unchanged. Thus the findings from Figs.
\ref{6new} and \ref{bild5} are robust against 'reasonable'
modifications of the in-medium transition rates.

The numerical abundancies  for $K^\pm$ mesons and antiprotons
relative to the average charged pion multiplicity $<\pi> = <\pi^+
+ \pi^->/2$ are displayed in Fig. \ref{bild4} for $Pb+Pb$ at 160
A$\cdot$GeV as a function of $A_{part}$. The comparison of the
calculations with the data from Ref. \cite{Na49b} indicates that
the dependence of the $K^\pm$ multiplicities on $A_{part}$ is
roughly met -- except a single point for $K^-$ at rather
peripheral reactions -- in line with the earlier analysis in Refs.
\cite{Cass99,Geiss}, which concentrated on central collisions in
this system.  However, the $\bar{p}/\pi$ ratio is  lower by about
a factor of 2 compared to the data of the NA49 Collaboration that
also include the feeddown from $\bar{\Lambda}$ and
$\bar{\Sigma}^0$  due to the weak interaction.  Since the latter
contribution is presently unknown one might either speculate that
the $\bar{\Lambda}+ \bar{\Sigma}^0$ abundancy is comparable to the
antiproton abundancy or that antiproton self energy effects might
be responsible for the experimental observation.

\subsubsection{Antiproton self energies}
As shown in Refs. \cite{Teis94,sibirtsev,Cass99,Ko96} the
production of antiprotons at SIS energies of 1.4--2.1 A$\cdot$GeV
as well as in $p+A$ reactions is described by adopting attractive
self energies for the antiprotons in the range of -100 to -150 MeV
at normal nuclear matter density $\rho_0$. Especially in $p+A$
reactions the backward production channels by a couple of mesons
are statistically irrelevant as can be easily checked by the
transport model described above. The question thus arises if such
'established' antiproton self energies for densities 1--3 $\rho_0$
might also be responsible for an enhancement of the $\bar{p}$
yield at SPS energies in the $Pb+Pb$ system.

To examine this possibility we show in Fig. \ref{bild6} the time
evolution of the baryon-density in a central cylinder of
transverse radius $R_T$ = 5 fm, that has moving boundaries with
the expanding hadronic system in longitudinal direction. The solid
line denotes the average density of 'formed' baryonic states
whereas the dashed line represents the net quark density
$\rho_q/3$ which merges with the baryon density in the later
expansion phase. Both densities have been evaluated in the cms
rapidity interval $|\Delta y|_{cm} \leq$ 1 in order to exclude
spectator nucleons and to gate on midrapidity physics. The
difference between the solid and dashed line in Fig. \ref{bild6}
has to be attributed to 'non-hadronic' states, which in the HSD
approach are quarks and diquarks (as well as their antiparticles)
that constitute the ends of 'strings' or continuum excitations of
the hadrons. As can be seen from Fig. \ref{bild6} the
'non-hadronic' phase lasts about 2.2 fm/c which is roughly the
diameter of the target ($2 R$) divided by the Lorentz-factor
$\gamma_{cm} \approx $ 9.3 plus the hadron formation time $\tau_F$
= 0.8 fm/c \cite{Cass99},
\be
\label{nonh} \tau_{nonhad.} \approx \frac{2 R}{\gamma_{cm}} +
\tau_F . \ee Then a mixed phase of 'partons' and 'formed' hadrons
comes up which practically ends around 6 fm/c where all continuum
excitations have merged to hadrons or hadronic resonant states.
Quite remarkably, the density of 'formed' baryons is about $2
\rho_0$ at the beginning of the pure hadronic expansion phase. Now
the $B\bar{B}$ annihilation rate (in Fig. \ref{bild5}) starts with
a maximum around 3 fm/c -- when the 'baryons' also start to
hadronize -- along with the
 meson fusion reactions that last up to about 8 fm/c. The
 characteristic time scale for the decrease of the annihilation/fusion rate is
 $\tau_{prod} \approx $1.6 fm/c for central $Pb+Pb$ reactions at
 the SPS as extracted from Fig. \ref{bild5} using an exponential ansatz.
  During the $B\bar{B}$ production
 phase by meson fusion from 3--8 fm/c
 the density of baryons drops from $\sim 2.5 \rho_0$ to $\rho_0$.
Employing  (\ref{estimate2}) the time scale for
 annihilation at these densities changes from 0.5 -- 1.2 fm/c,
 which is considerably shorter than the characteristic production
 time scale of 1.6 fm/c. The approximate chemical equilibration between mesons and
 antibaryons thus is no surprise according to these simple 'classical'
 estimates.
 On the other hand, the densities of 2.5 $\rho_0$ -- $\rho_0$ are
 very similar to the densities probed in heavy-ion
 reactions at SIS energies from 1--2 A$\cdot$GeV \cite{Ko96,Cassing90}.

In order to investigate if $\bar{p}$ self energies might be
responsible for the experimental antiproton yield, a scalar
attractive $\bar{p}$ potential of the form
\be
\label{pot} U_{\bar{p}}(\rho) = - \alpha \frac{\rho}{\rho_0} \ee
is assumed with $\alpha$ = 100 MeV, where $\rho$ denotes the
density of 'formed' baryons. This amounts to produce and propagate
antiprotons with a density-dependent mass $M^* = M_0 +U(\rho)$.
For more technical details the reader is referred to Refs.
\cite{Teis94,sibirtsev}. Note, that when introducing self energies
for particles by averaging the latter over many events one no
longer performs {\it microcanonical} simulations since the field
energy is shared between different events. However, quark flavor
conservation still holds exactly in each event such that the
physical system corresponds to a {\it canonical} ensemble.

The numerical results for the $\bar{p}/<\pi>$ ratio in $Pb + Pb$
collisions at 160 A$\cdot$GeV are displayed in Fig. \ref{bild7} by
the solid line in comparison to the cascade calculation (dashed
line) and the experimental data from NA49 \cite{Na49b} (full
triangles). It is seen that with increasing centrality or
$A_{part}$ there is a slight enhancement of the $\bar{p}$ yield
for the potential (\ref{pot}), which vanishes for very peripheral
reactions where the average baryon density is very small. The
experimental data, however, are still underestimated
significantly. This also holds true when accounting for the 20\%
overprediction of pions in the transport approach (cf. Fig. 5).

\subsubsection{Extrapolation for antihyperon production}
In view of the approximate chemical equilibrium achieved for the
$u,d$ quark sector between mesons and antibaryons -- even for
reduced annihilation cross sections -- one may proceed with
speculations on the strangeness ($s\bar{s}$-quark) sector, where
no experimental data on the annihilation cross sections are
available. However, in case of similar transition matrix elements
squared ($\approx$ 5 fm$^2$) the $Y \bar{N}$ and $\bar{Y} N$
channels, where $Y$ stands for the hyperons ($\Lambda, \Sigma,
\Sigma^*$), will achieve chemical equilibrium with the 3 meson
system as before, however, with a $\pi$ or $\rho, \omega$
exchanged by a $K$ or $K^*,\phi$, respectively (cf. Ref.
\cite{Carsten} for the case of 5 meson reaction channels). A
further step then consists in replacing another pion or $\rho,
\omega$ by a $K$ or $K^*,\phi$ which will bring the $\Xi, \Xi^*$
and $\bar{\Xi}, \bar{\Xi}^*$ in chemical equilibrium with the
$SU(3)_{flavor}$ meson system. Moreover, in case of flavor
rearrangement reactions with 3 strange mesons $K, K^*, \phi$ (or
any combinations of those) the $\Omega, \bar{\Omega}$ system might
also achieve chemical equilibrium. This will imply for the
antibaryon to baryon ratios:
\be
\label{chemr} \frac{\bar{p}}{p} \approx \frac{K^-}{K^+} \
\frac{\bar{\Lambda}}{\Lambda} \approx \left ( \frac{K^-}{K^+}
\right )^2 \ \frac{\bar{\Xi}}{\Xi} \approx \left ( \frac{K^-}{K^+}
\right )^3 \ \frac{\bar{\Omega}}{\Omega} . \ee The relations
(\ref{chemr}) are easily obtained for a {\it grand canonical}
ensemble at 'high' temperature $T$ where Fermi and Bose
distributions coincide with the Boltzmann distribution. In this
limit the ratios of particles to antiparticles (apart from the
temperature $T$) only depend on the chemical potential $\mu_q$ for
light quarks and $\mu_s$ for strange quarks, i.e. $$
\frac{\bar{p}}{p} = \frac{\exp(-(E_{\bar{p}} +3
\mu_q)/T)}{\exp-(E_p -3 \mu_q)/T)} = \exp(-6 \mu_q/T), $$ $$
\frac{\bar{K}}{K} = \frac{\exp(-(E_{\bar{K}} - \mu_s +
\mu_q)/T)}{\exp(-(E_K + \mu_s - \mu_q)/T)} = \exp(2(\mu_s-
\mu_q)/T), \ $$ $$ \frac{\bar{\Lambda}}{\Lambda} =
\frac{\exp(-(E_{\bar{\Lambda}} + \mu_s + 2
\mu_q)/T)}{\exp(-(E_\Lambda - \mu_s - 2 \mu_q)/T)} = \exp(-(2
\mu_s+ 4 \mu_q)/T) \ $$ $$ \frac{\bar{\Xi}}{\Xi} =
\frac{\exp(-(E_{\bar{\Xi}} + 2 \mu_s + \mu_q)/T)}{\exp(-(E_\Xi - 2
\mu_s -  \mu_q)/T)} = \exp(-(4 \mu_s+ 2 \mu_q)/T) \ $$
\be
\frac{\bar{\Omega}}{\Omega} = \frac{\exp(-(E_{\bar{\Omega}} + 3
\mu_s )/T)}{\exp(-(E_\Omega - 3 \mu_s )/T)} = \exp(-6 \mu_s/T), \
\ee
 where $E_X, E_{\bar{X}}$ denote the energies of particles and
antiparticles, respectively, that are the same when neglecting
self energies. Note, that the relations (\ref{chemr}) also result
from the quark condensation model in Refs.
\cite{Zimanyi1,Zimanyi2}, however, involve a slightly different
physical picture. In the latter approach the mesons, baryons and
antibaryons emerge from an equilibrium QGP state by condensation
under the constraint of quark flavor conservation. In the
microcanonical transport approach discussed here, the relations
(\ref{chemr}) come about due to the strong annihilation of
antibaryons with baryons and the backward flavor rearrangement
channels by detailed balance, i.e. by purely hadronic reaction
channels in the expansion phase of the system.

It should be pointed out, that the canonical statistical approach
of Ref. \cite{Redlich2} -- involving an additional parameter of
dimension fm$^3$ -- well describes the strange and
multi-strange baryon and antibaryon abundancies in $Pb+Pb$
collisions at SPS energies. Thus the concept of chemical
equilibrium also on the strangeness sector $S$= $\pm 1,\pm 2,\pm 3$ appears
compatible with the experimental observations. However, it is
argued here that the relations (\ref{chemr}) are a consequence of
the strong flavor rearrangement reactions between formed hadrons and
do not signal the presence of a QGP state.

Assuming chemical equilibrium also for baryons and antibaryons
with strangeness then (by Eq. (\ref{chemr})) the $\bar{\Lambda}$
multiplicity is related to the antiproton multiplicity as
\be
\label{antihyp} \bar{\Lambda} \approx \left ( \frac{K^+}{K^-}
\times \frac{\Lambda}{p}\right ) \ \bar{p}. \ee All quantities on
the r.h.s. of (\ref{antihyp}) are known from the transport
calculation, however, only the interacting number of protons have
to be counted in this case since proton spectators have to be
excluded in this balance. A rather save way is to take into
account only particles at midrapidity for $|\Delta y| \leq$ 1,
which then excludes spectator baryons. Thus counting only
particles at midrapidity for the ratio in the brackets in
(\ref{antihyp}) the $\bar{\Lambda}$ abundancy is entirely
determined by particle ratios that have sufficient statistics in
the transport calculation. The resulting $(\bar{p} +
\bar{Y})/<\pi>$ ratio for $Pb + Pb$ at 160 A$\cdot$GeV is shown in
Fig. \ref{bild7} by the open circles with error bars that are due
to particle statistics in the transport calculation. In this case
the experimental multiplicity is reproduced rather well from
peripheral to central collisions (within the error bars)
suggesting the ratio $\bar{\Lambda}/\bar{p} \approx $ 1 for a wide
range of impact parameters. However, a precise experimental
separation of antiprotons from antihyperons  will be necessary to clarify the present
ambiguities.

\subsection{AGS energies}
The experimental information on antibaryon production at AGS
energies is rather scarce and limited to specific rapidity
intervals. Since antibaryon yields are also reduced substantially
as compared to SPS energies this imposes severe constraints on the
statistics in nonperturbative transport calculations. Thus, before
addressing any comparison to experimental data, we show in Fig.
\ref{bild8} the density of 'formed' baryons in a central expanding
cylinder of transverse radius $R_T$ = 5 fm for $Au + Au$ at 11
A$\cdot$GeV in comparison to the net quark density $\rho_q/3$
(dashed line). When comparing to Fig. \ref{bild6} for $Pb + Pb$ at
160 A$\cdot$GeV roughly the same maximum net quark density is
found for $|\Delta y| \leq$ 1, however, the nonhadronic phase
characterized by (\ref{nonh}), i.e. $\tau_{nonhad.} \approx 6.2$
fm/c, lasts much longer due to the lower Lorentz $\gamma$-factor
$\gamma_{cm} \approx 2.6$. Furthermore, the density of 'formed'
baryons is lower than at SPS energies since most of these hadronic
states rescatter again on baryons in the medium since the
formation time $\tau_F$ = 0.8 fm/c is small compared to the
reaction time roughly given by $2 R/\gamma_{cm}$, where $R$
denotes the radius of the $Au$ nucleus. Thus 'formed' baryons are
reexcited to strings for a couple times during the nucleus-nucleus
collision. This phenomenon has been addressed as 'string matter'
in Ref. \cite{Sahu00} and should not be interpreted as a state of
quark-gluon plasma (QGP).

A comparison of the HSD transport calculations with the
experimental data of the E866 Collaboration at midrapidity
\cite{AGSall} is presented in Fig. \ref{bild9} as a function of
the participating protons $N_{pp}$ for $Au + Au$ at 11.6
A$\cdot$GeV/c. Whereas the proton to $\pi^+$ ratio is rather well
described as a function of centrality (lower left part) the
$K^+/\pi^+$ and $K^-/\pi^+$ ratios are underestimated
systematically for all centralities when discarding self energies
for the strange hadrons (cf. Refs. \cite{Cass99,Geiss,Cass00a}).
The calculated $K^+/K^-$ ratio from the HSD calculation is 5$\pm
0.3$ for all centralities rather well in line with the
experimental observation at midrapidity. Since strangeness
conservation is exactly fulfilled in the calculations this
demonstrates that the net production of $s\bar{s}$ quarks by
'hard' baryon-baryon, baryon-meson and meson-meson reactions is
underestimated in the transport approach as discussed in more
detail in Refs. \cite{Cass99,Geiss}. On the other hand the
antiproton to $\pi^+$ ratio (lower right part) is compatible with
the experimental ratios within the error bars indicating an
approximately constant value of $\sim 2.5-3 \times 10^{-4}$. This
roughly constant $\bar{p}/\pi^+$ ratio is a consequence of an
approximate chemical equilibration as demonstrated in Fig.
\ref{bild10} for a central collision of $Au+Au$ at 11.6
A$\cdot$GeV/c. Here the solid histogram corresponds to the
annihilation rate of antibaryons whereas the dashed histogram
stands for the backward 3 meson fuse rate. Though the statistics
are very limited in this case a net absorption of antibaryons,
i.e. the difference in the time integrals of the annihilation rate
and 3 meson production rate, is still present in the calculations
which amounts to about 20\% for central reactions and about 30\%
for very peripheral reactions of the total number of antibaryons
produced in baryon-baryon  or meson-baryon reactions. This
relative net absorption, however, can only be extracted from
transport calculations and is not a measurable quantity
experimentally.

The E877 Collaboration, furthermore, has observed a sizeable
anti-flow of $\bar{p}$'s in $Au + Au$ reactions at 11.6
A$\cdot$GeV/c  \cite{E877} which either indicates a strong
absorption of antiprotons on baryons or the current of comoving
mesons, that fuse to $B\bar{B}$ pairs, and are anticorrelated to
the proton current themselves. The present transport calculations
reproduce these correlations, however, suffer from large
statistical error bars (similar in size to those of the data) such
that an explicit comparison is discarded in this work.

As mentioned in the introduction, a high ratio of $\bar{\Lambda}$
to $\bar{p}$ of $3.6^{+4.7}_{-1.8}$ has been reported by the E917
Collaboration \cite{AGSnew} for central collisions of $Au + Au$ at
11.7 A$\cdot$GeV/c that is not understood so far. These data, furthermore,
are in a qualitative agreement with the measurements from the E864/E878
Collaboration \cite{E864n}. To estimate the
$\bar{\Lambda}/\bar{p}$ ratio within the present transport
approach as a function of the centrality of the collision we employ
again the relation (\ref{antihyp})  and invoke
the particle multiplicities at midrapidity $|\Delta y_{cm}| \leq
1$. The results of these calculations are displayed in Fig.
\ref{bild11} for $Au+Au$ at 11.6 A$\cdot$GeV/c as a function of
the number of participating protons indicating a steady rise of
the ratio with centrality. The hatched area in Fig. \ref{bild11}
demonstrates the uncertainty in the $\bar{\Lambda}/\bar{p}$ ratio
due to the limited statistics. However, the calculations for the
most central collisions do not suggest ratios above 1.4 which is
still slightly out of the range of the number quoted in Ref.
\cite{AGSnew} of $3.6^{+4.7}_{-1.8}$. Thus either new theoretical
concepts and/or much refined data are necessary to unravel this
puzzle.

\section{Summary}
In this work the conventional transport approach for two-body
induced reactions has been extended on the formal level to n-body
$n \leftrightarrow m$ reaction channels employing the principles
of detailed balance. As a specific example of current interest the
baryon-antibaryon annihilation problem in relativistic
nucleus-nucleus collisions has been addressed at AGS and SPS
energies where the meson densities are comparable (at the AGS) or
much larger than the baryon densities (at the SPS) such that
multiple meson fusion reactions as suggested in Refs.
\cite{ko1,Rapp,Carsten} become very probable.

\begin{figure}[h]
\centerline{\psfig{file=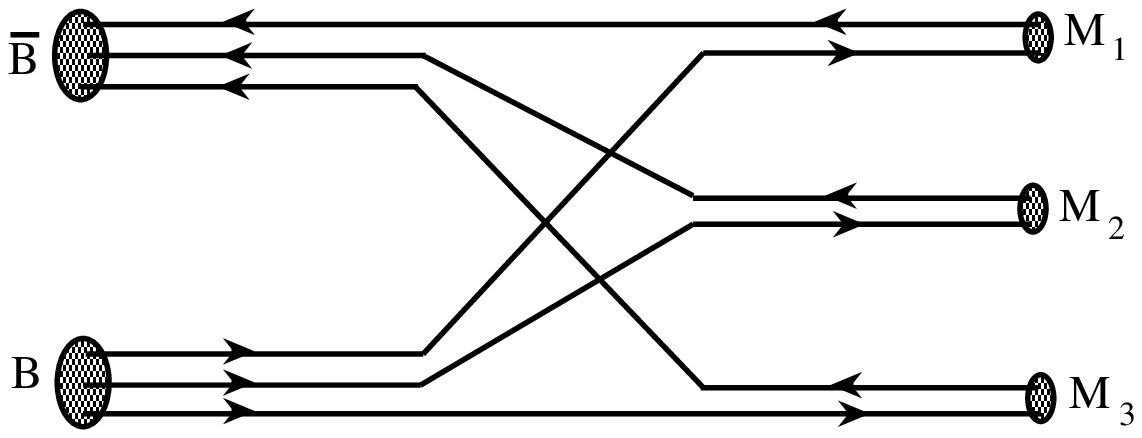,height=10cm}}
\caption{Illustration of the flavor rearrangement model for
$B\bar{B}$ annihilation to 3 mesons and vice versa. The mesons
$M_i$ may be either pseudo-scalar or vector mesons, respectively.}
\label{bild1}
\end{figure}

\begin{figure}[h]
\centerline{\psfig{file=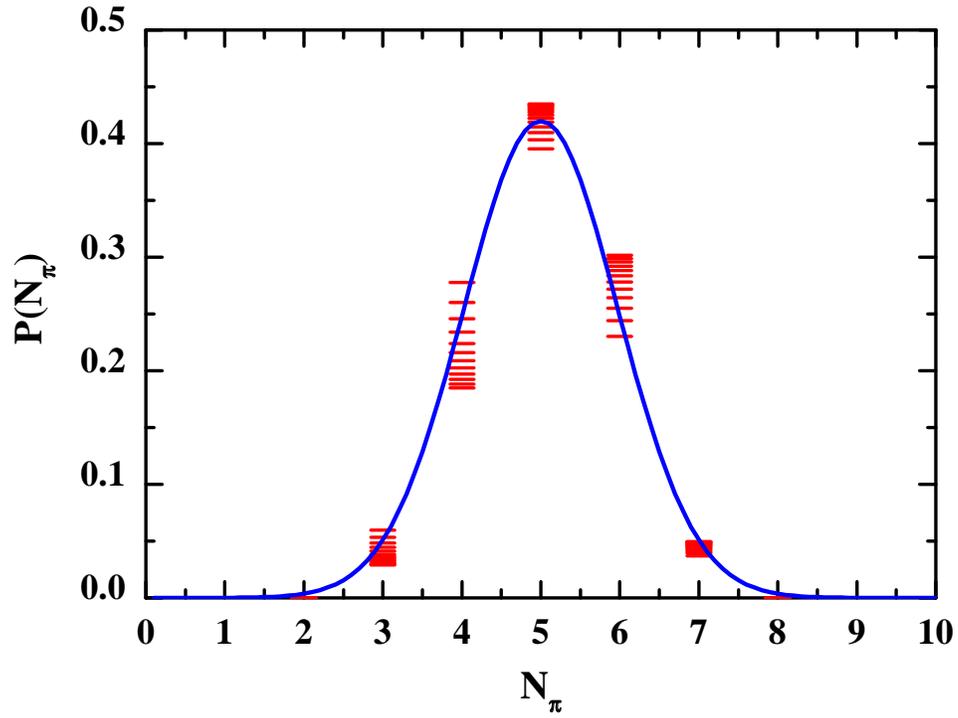,width=15cm}}
\vspace*{-5cm}
\caption{The distribution in the final number of pions $P(N_\pi)$
for $p\bar{p}$ annihilation at invariant energies 2.3 GeV $\leq
\sqrt{s} \leq$ 4 GeV (short lines). The solid line is the gaussian
parametrization (\protect\ref{pion5}) that is fitted to the experimental data.}
\label{bild2}
\end{figure}

\begin{figure}[h]
\centerline{\psfig{file=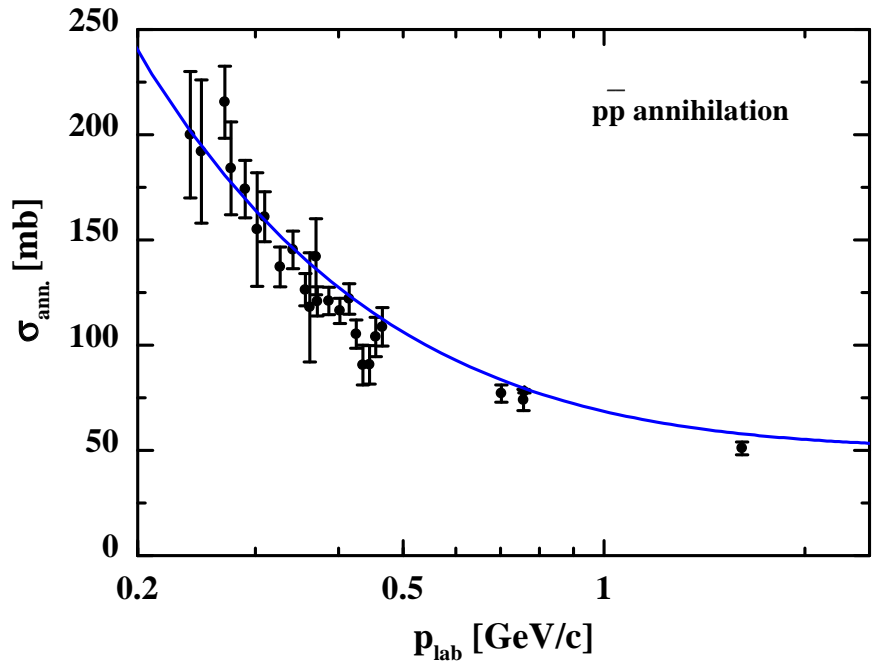,height=20cm}} \caption{The
annihilation cross section $\sigma_{ann.}$ for the $p\bar{p}$
reaction as a function of the laboratory momentum $p_{lab}$ from
Ref. \protect\cite{LB} in comparison to the approximation
(\ref{appann}) (solid line). }
 \label{bild1b}
\end{figure}

\begin{figure}[h]
\centerline{\psfig{file=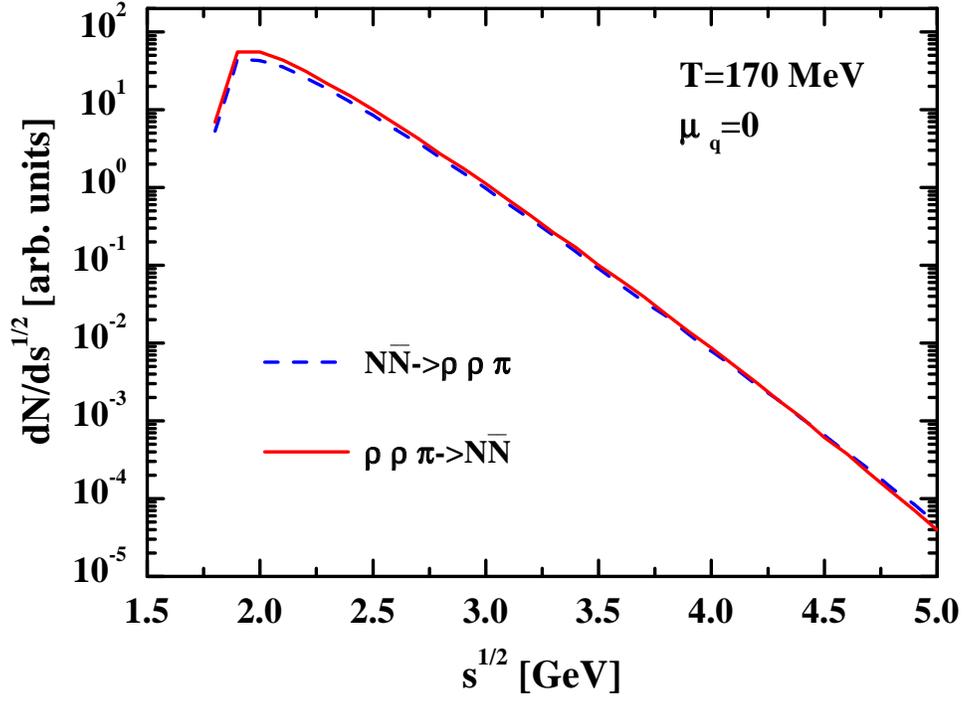,width=15cm}}
\vspace*{-5cm}
\caption{The number of $N\bar{N} \rightarrow \rho \rho \pi$ reactions as a function
of the invariant energy $\sqrt{s}$ for a system in thermal and chemical equilibrium
at temperature $T$ = 170 MeV and $\mu_q$ = 0. The solid line denotes the differential number
in the backward ($\rho \rho \pi$) collisions, respectively.}
\label{bild3}
\end{figure}

In order to employ the many-body detailed balance relations (as
addressed in Section 2) a simple phase-space model for antibaryon
annihilation has been presented that is based on flavor
rearrangement channels to pseudoscalar and vector mesons (cf. Fig.
\ref{bild1}). Furthermore, a suitable covariant scheme for the
calculation of such multi-particle reactions has been presented in
Section 3 which is a straight forward extension of the concept
proposed by Lang et al. in Ref. \cite{Lang93}. The method and its
implementation in the HSD transport approach \cite{Cass99,Ehehalt}
has been tested for a homogenuous system of nonstrange hadrons in
thermal and chemical equilibrium (cf. Fig. \ref{bild3}).

Actual transport calculations have been performed for $Pb + Pb$ at
160 A$\cdot$GeV and $Au+Au$ at 11.6 A$\cdot$GeV/c, i.e. the most
prominant reactions at the SPS and the AGS, where a couple of
experimental data are available to control the dynamics. It is
found that at both energies the meson fusion reactions are by far
the most dominant production channel for the final antibaryons
(seen experimentally) and that the antibaryons and baryons -- at
least at midrapidity -- come close to local chemical equilibrium
with the mesons. As a consequence the $\bar{p}/\pi$ ratio is
practically independent on the centrality of the collision, i.e.
as a function of $A_{part}$, in line with the experimental
observation at both energies.

On the other hand, the approximate
chemical equilibration allows to perform rather reliable
extrapolations for the strange and multistrange antibaryon
abundancies on the basis of particle ratios (cf. (\ref{chemr})),
that follow from chemical equilibrium for the mesons and antibaryons.
The latter ratios can be calculated with better statistics in the
nonperturbative approach than the direct strange antibaryon abundancies.
Here a roughly constant
$\bar{\Lambda}/\bar{p}$ ratio of $\approx$ 1 is predicted for
semi-peripheral to central collisions of $Pb+Pb$ at the SPS that
will be controlled soon by experimental data from the NA49
Collaboration. Moreover, a separation of antiprotons from
antihyperons  will also allow to investigate the question of
antiproton self energies that enhance the $\bar{p}/\pi$ ratio with
centrality when adopting attractive scalar potentials in line with
the analysis performed at SIS energies of $\sim$ 2 A$\cdot$GeV
\cite{sibirtsev}. At AGS energies of 11.6 A$\cdot$GeV/c the
situation is less clear due to the rather low statistics for
antibaryons, both theoretically and experimentally. Recall that
the $\bar{p}/\pi^+$ ratio is only about $2.5-3 \times 10^{-4}$ at
this energy. The HSD transport calculations for the most central
collisions of $Au+Au$ at the AGS -- assuming (\ref{chemr}) to hold --
give an upper limit of 1.4 for
the $\bar{\Lambda}/\bar{p}$ ratio, which is still slightly out of
the range of the number quoted by the E917 Collaboration
\cite{AGSnew} of $3.6^{+4.7}_{-1.8}$ or related results from Ref. \cite{E864n}.
Thus either new theoretical
concepts and/or much refined data will be necessary to unravel the
antihyperon puzzle at  AGS energies.

\vspace{0.5cm} The author likes to acknowledge continuous and
valuable discussions with C. Greiner. Furthermore, he is indepted
to E. L. Bratkovskaya and C. Greiner for critical suggestions and
a careful reading of the manuscript.

\newpage

\begin{figure}[h]
\centerline{\psfig{file=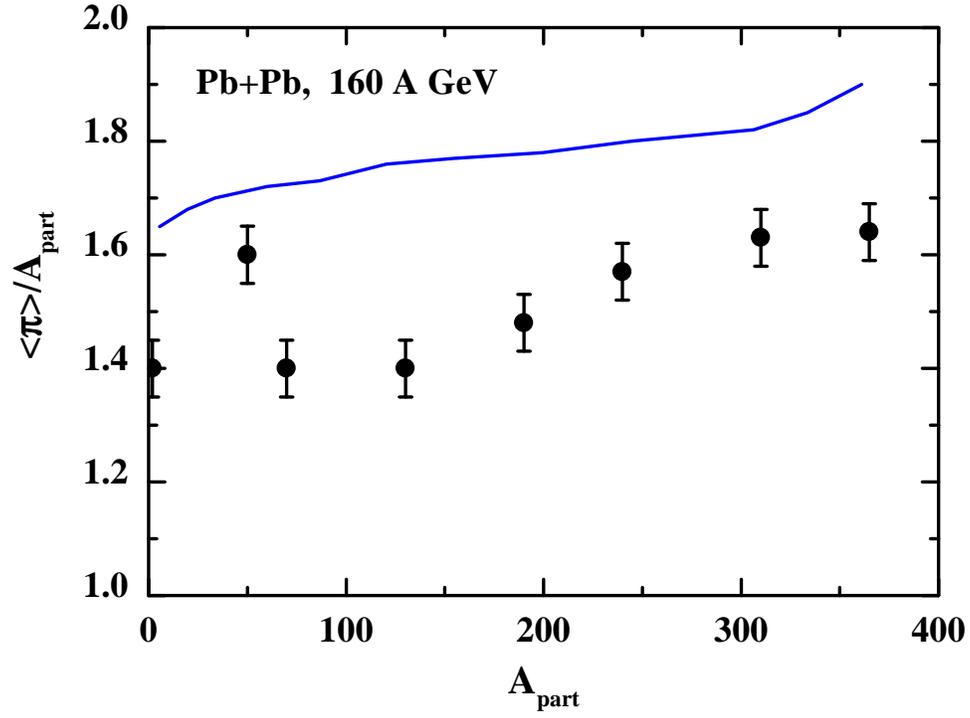,width=15cm}}
\vspace*{-5cm}
\caption{The calculated average number of charged pions $<\pi>$ divided by $A_{part}$
as a function of centrality for $Pb+Pb$ at 160 A$\cdot$GeV in comparison to the data from Ref.
\cite{Na49b}.}
\label{fig4}
\end{figure}
\begin{figure}[h]
\centerline{\psfig{file=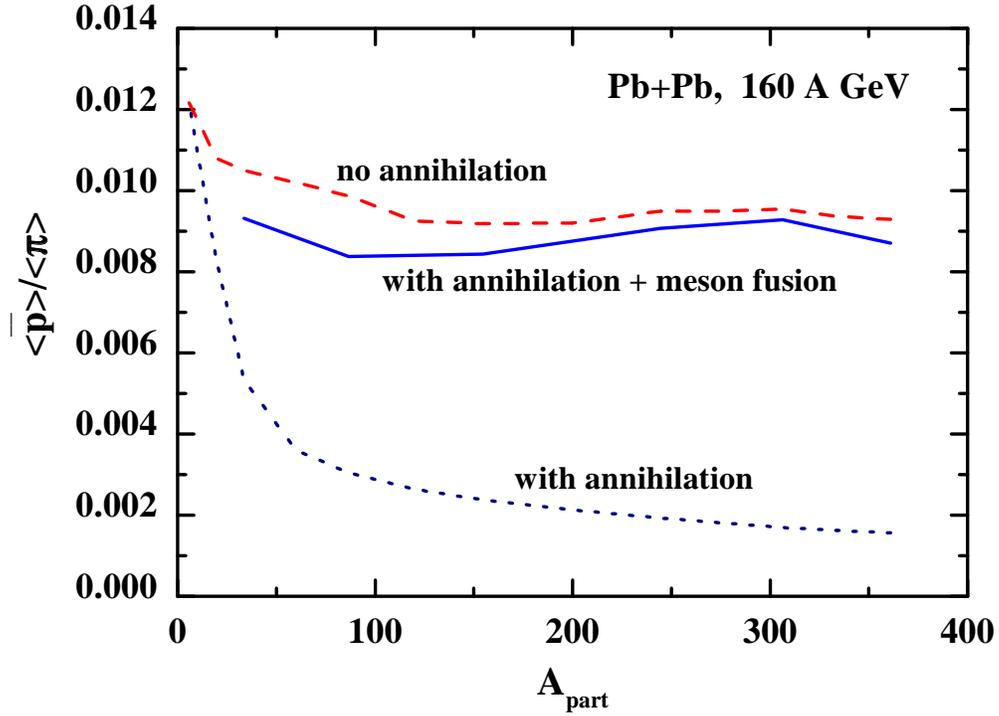,width=15cm}}
\vspace*{-5cm}
\caption{The ratio of antiprotons to the average number of charged pions $<\pi>$
as a function of centrality for $Pb+Pb$ at 160 A$\cdot$GeV within different approximations.
The dashed line reflects calculations without annihilation of
antibaryons, the dotted line includes the annihilation channels
while the solid line stands for the calculations with both, the
annihilation and meson fusion channels.}
\label{6new}
\end{figure}
\begin{figure}[h]
\centerline{\psfig{file=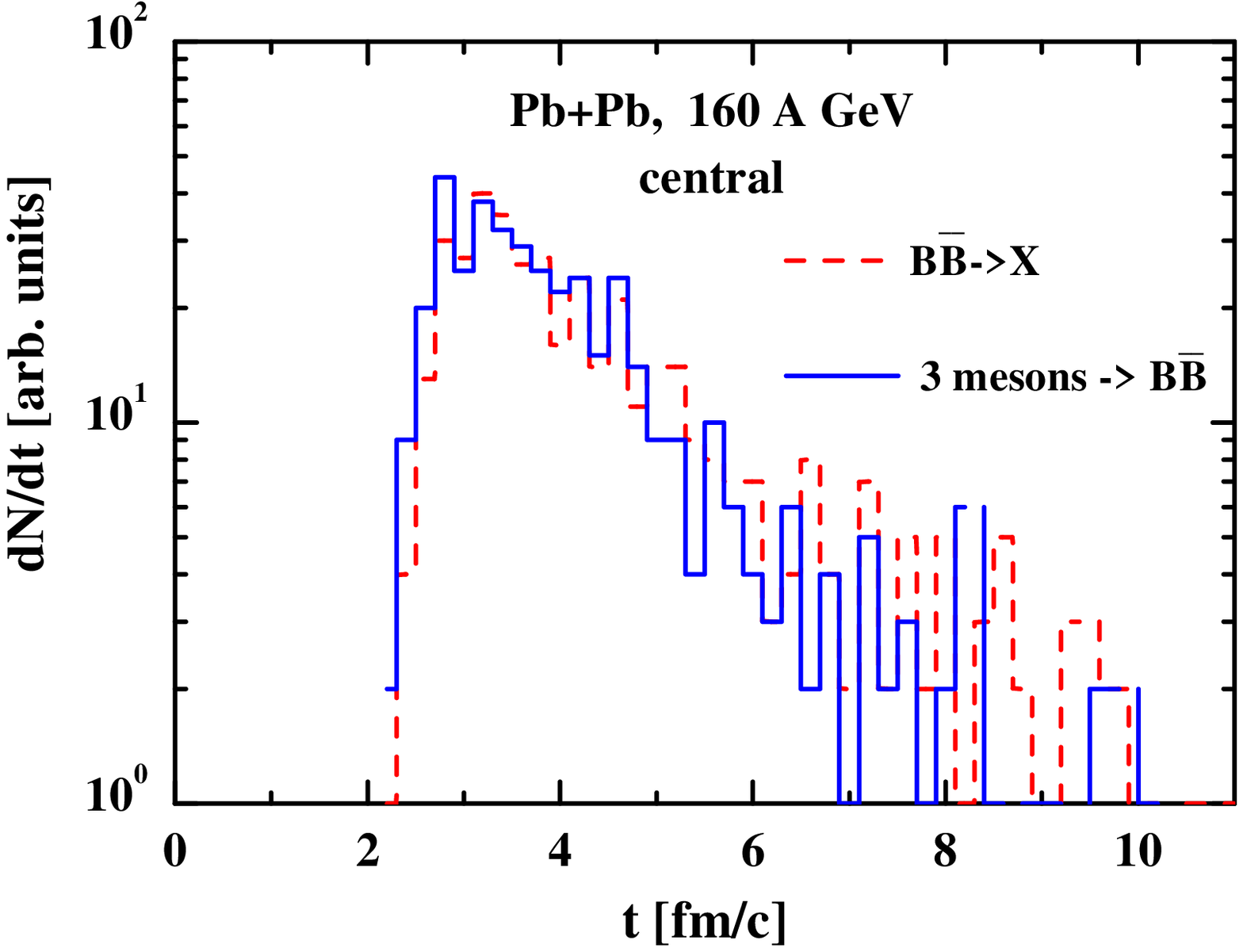,width=15cm}}
\vspace*{-5cm}
\caption{The annihilation rate $B\bar{B} \rightarrow 3 mesons$ (dashed histogram)
for a central $Pb+Pb$ collision at 160 A$\cdot$GeV as a function
of time in comparison to the backward reaction rate (solid
histogram) within the HSD transport approach.}
\label{bild5}
\end{figure}
\begin{figure}[h]
\centerline{\psfig{file=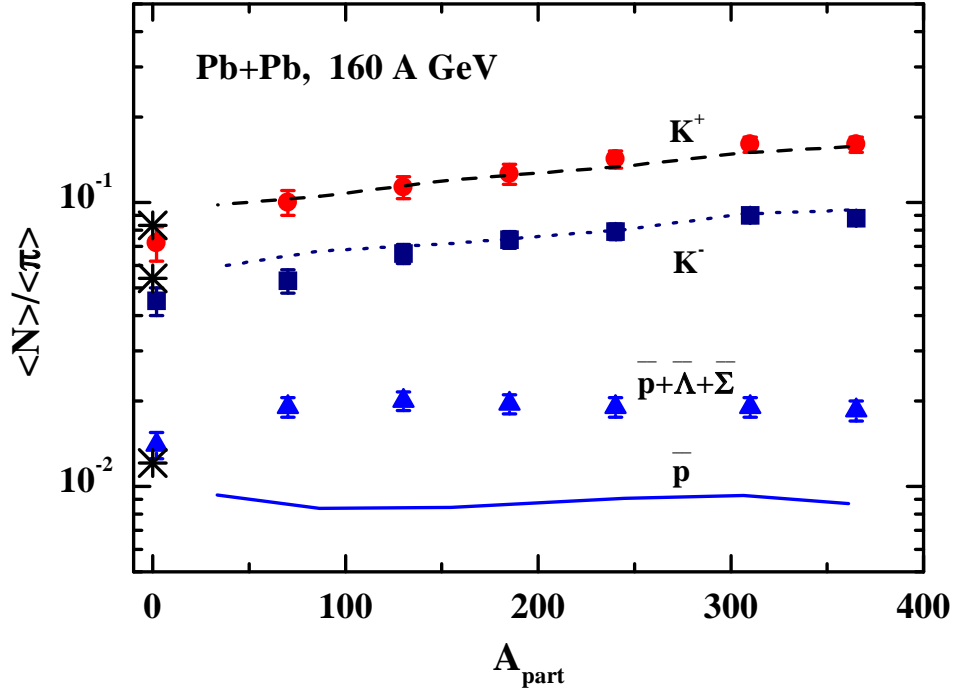,width=15cm}} \vspace*{-5cm}
\caption{The calculated $K^+/\pi$ (dashed), $K^-/\pi$ (dotted) and
$\bar{p}/\pi$ (solid) ratio for $Pb + Pb$ at 160 A$\cdot$GeV as a
function of the number of participating nucleons $A_{part}$. The
experimental data are taken from Ref. \cite{Na49b}; the latter
data for antiprotons include also the feeddown from $\bar{\Lambda}$ and
$\bar{\Sigma}^0$, respectively. The stars for $A_{part}$ = 2 denote the
transport results for $K^+/\pi$, $K^-/\pi$ and $\bar{p}/\pi$ ratios in case of
$NN$ reactions at $T_{lab}$ = 160 GeV.} \label{bild4}
\end{figure}
\begin{figure}[h]
\centerline{\psfig{file=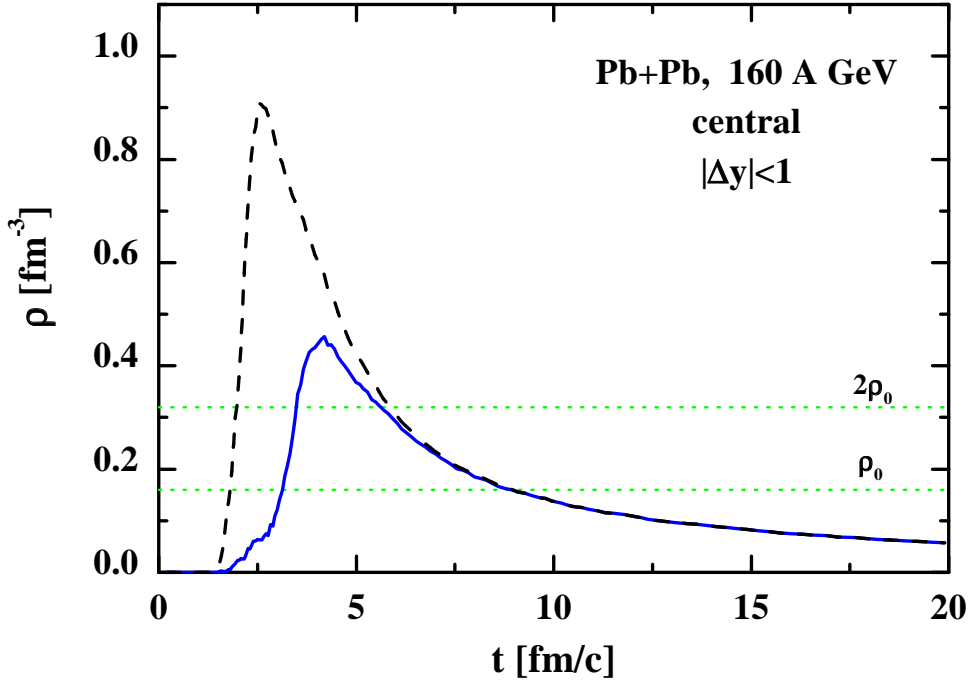,width=15cm}}
\vspace*{-5cm}
\caption{The baryon density in a central expanding cylinder (see text) for $|\Delta y|_{cm} \leq$
 1 in case of a central collision of $Pb+Pb$ at 160 A$\cdot$GeV in the HSD approach as a function of time.
 The solid line shows the density of 'formed' baryonic states
 while the dashed line stands for the net quark density
 $\rho_q/3$, while merges with the baryon density in the expansion
 phase. The dotted lines at $\rho_0$ and $2 \rho_0$ are drawn to guide the eye.}
\label{bild6}
\end{figure}
\begin{figure}[h]
\centerline{\psfig{file=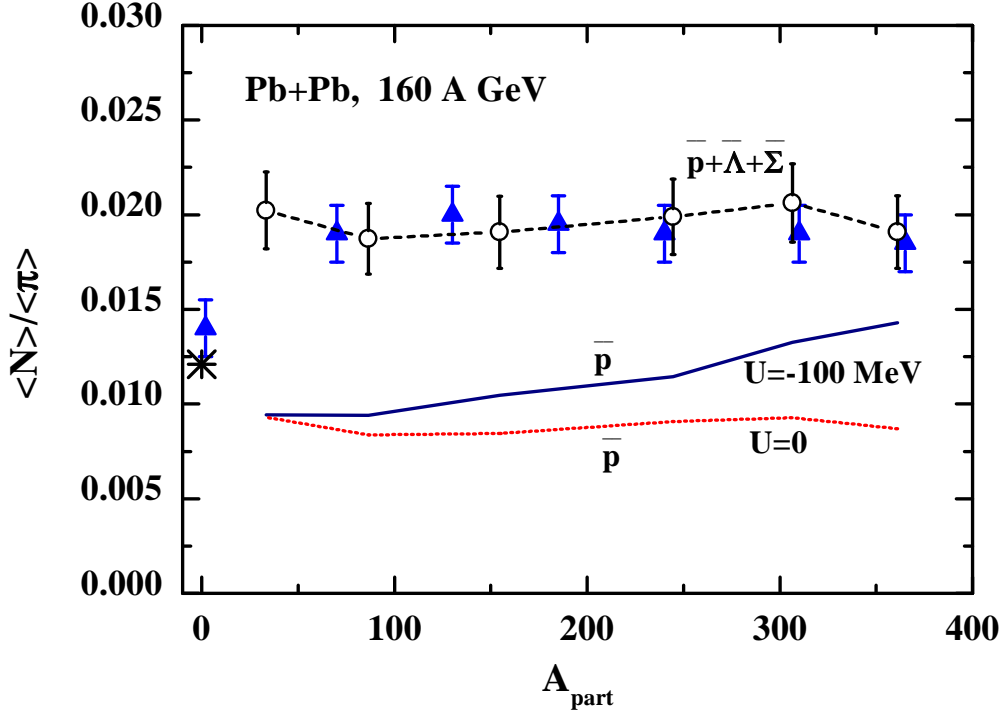,width=15cm}} \vspace*{-5cm}
\caption{The antiproton to charged pion ratio (see text) as a
function of the number of participating nucleons $A_{part}$ for
$Pb+Pb$ at 160 A$\cdot$GeV. The dotted line denoted by $U=0$
corresponds to the cascade result (cf. Fig. \ref{bild4}) while the
solid line denoted by $U= - 100$ MeV corresponds to a calculation
with the attractive scalar antiproton potential (\ref{pot}) of
-100 MeV at baryon density $\rho_0$. The full triangles represent
the data from Ref. \cite{Na49b} while the open circles correspond
to a cascade calculation including the feeddown from antihyperons
according to Eq. (\ref{antihyp}). The errorbars are due to the
limited statistics in the transport calculation. The star for
$A_{part}$ = 2 denotes $\bar{p}/\pi$ ratio for $NN$ collisions at
$T_{lab}$ = 160 GeV. } \label{bild7}
\end{figure}
\begin{figure}[h]
\centerline{\psfig{file=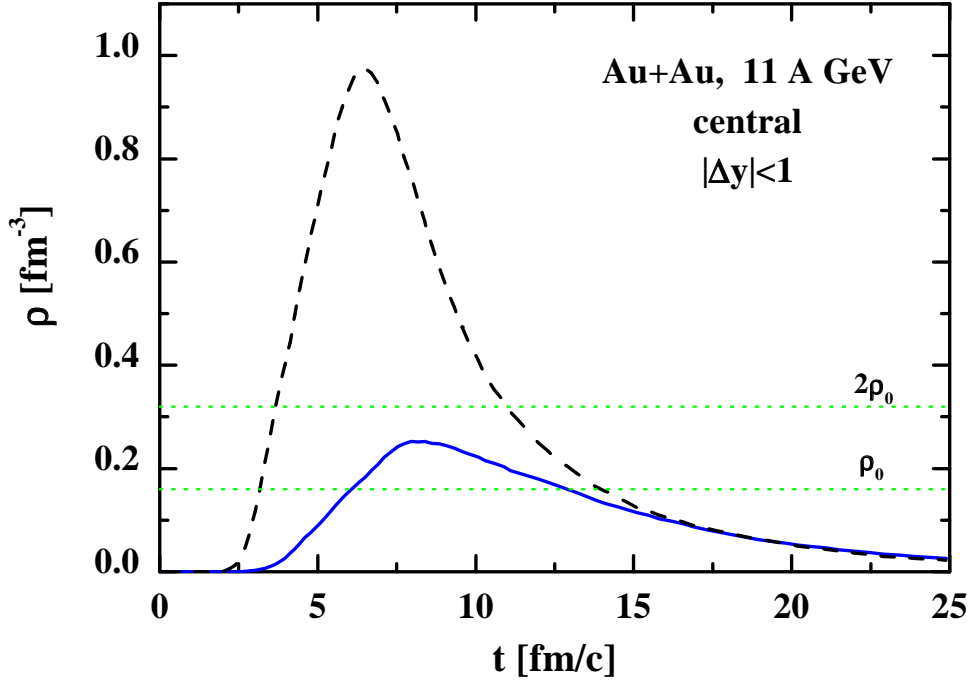,width=15cm}} \vspace*{-5cm}
\caption{The baryon density in a central expanding cylinder (see
text) for $|\Delta y|_{cm} \leq$
 1 in case of a central collision of $Au+Au$ at 11.6 A$\cdot$GeV/c in the HSD approach
 as a function of time.
 The solid line shows the density of 'formed' baryonic states
 while the dashed line stands for the net quark density
 $\rho_q/3$, while merges with the baryon density in the expansion
 phase. }
\label{bild8}
\end{figure}
\begin{figure}[h]
\centerline{\psfig{file=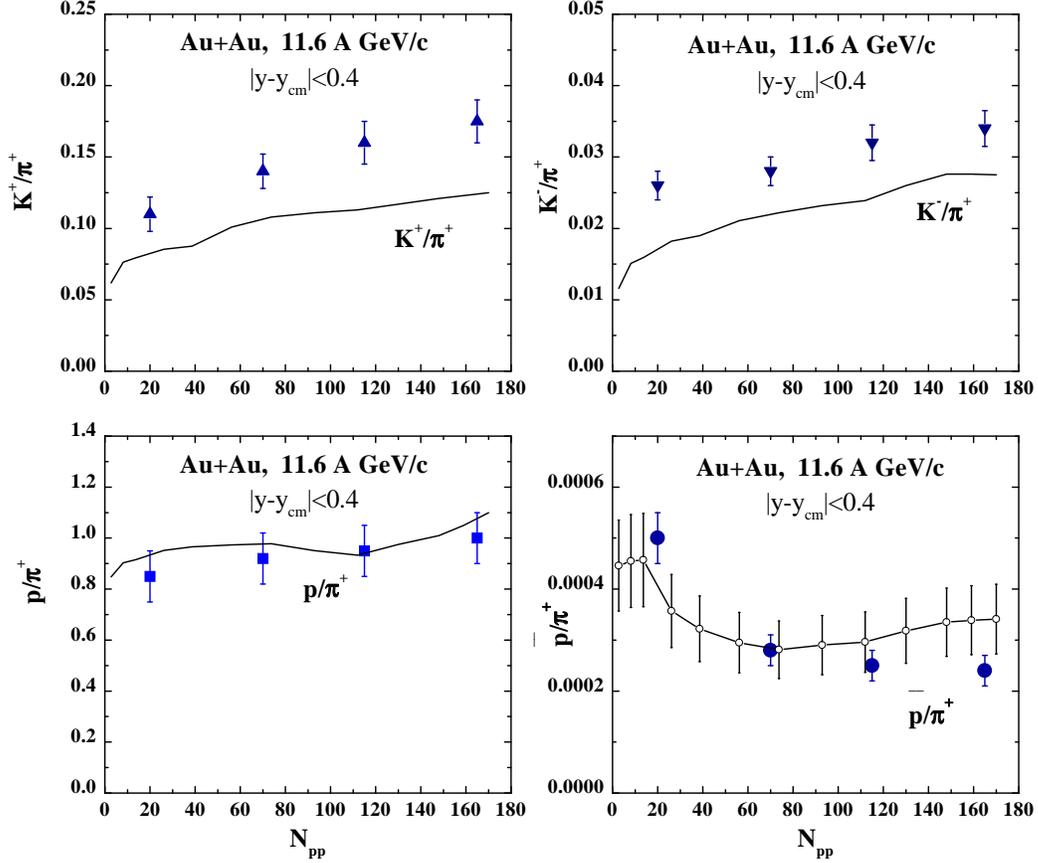,width=15cm}} \vspace*{-5cm}
\caption{The $K^+/\pi^+$, $K^-/\pi^+$, $p/\pi^+$ and
$\bar{p}/\pi^+$ ratio at midrapidity  for $Au + Au$  at 11.6
A$\cdot$GeV/c as a function of the number of participating protons
$N_{pp}$. The solid line correspond to the HSD transport
calculation in the cascade mode for mesons and antibaryons (open
circles)  while the experimental data (full symbols) are taken
from \cite{AGSall}.} \label{bild9}
\end{figure}
\begin{figure}[h]
\centerline{\psfig{file=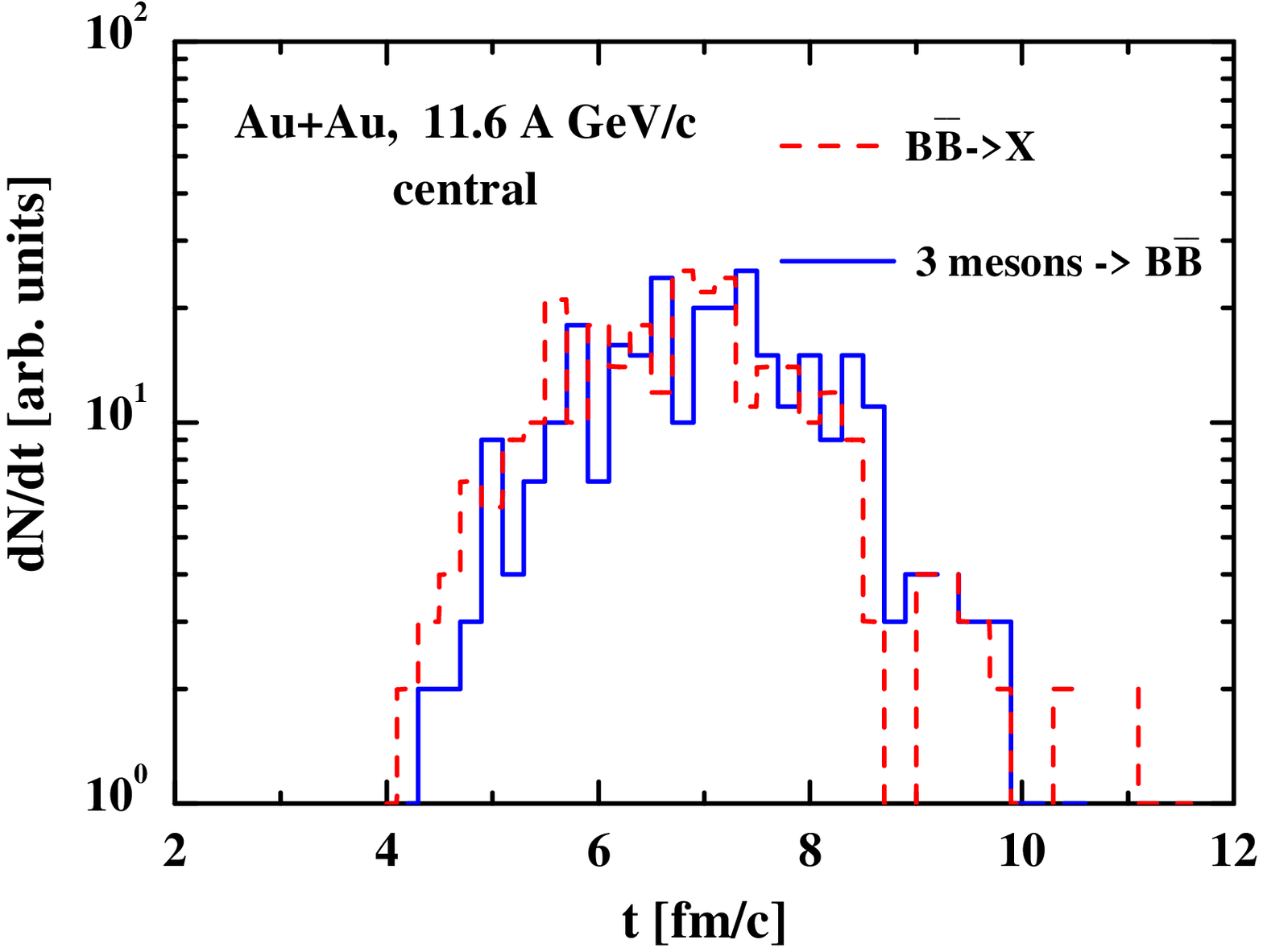,width=15cm}} \vspace*{-5cm}
\caption{The annihilation rate $B\bar{B} \rightarrow 3 mesons$
(dashed histogram) for a central $Au+Au$ collision at 11.6
A$\cdot$GeV/c as a function of time in comparison to the backward
reaction rate (solid histogram) within the HSD transport
approach.} \label{bild10}
\end{figure}
\begin{figure}[h]
\centerline{\psfig{file=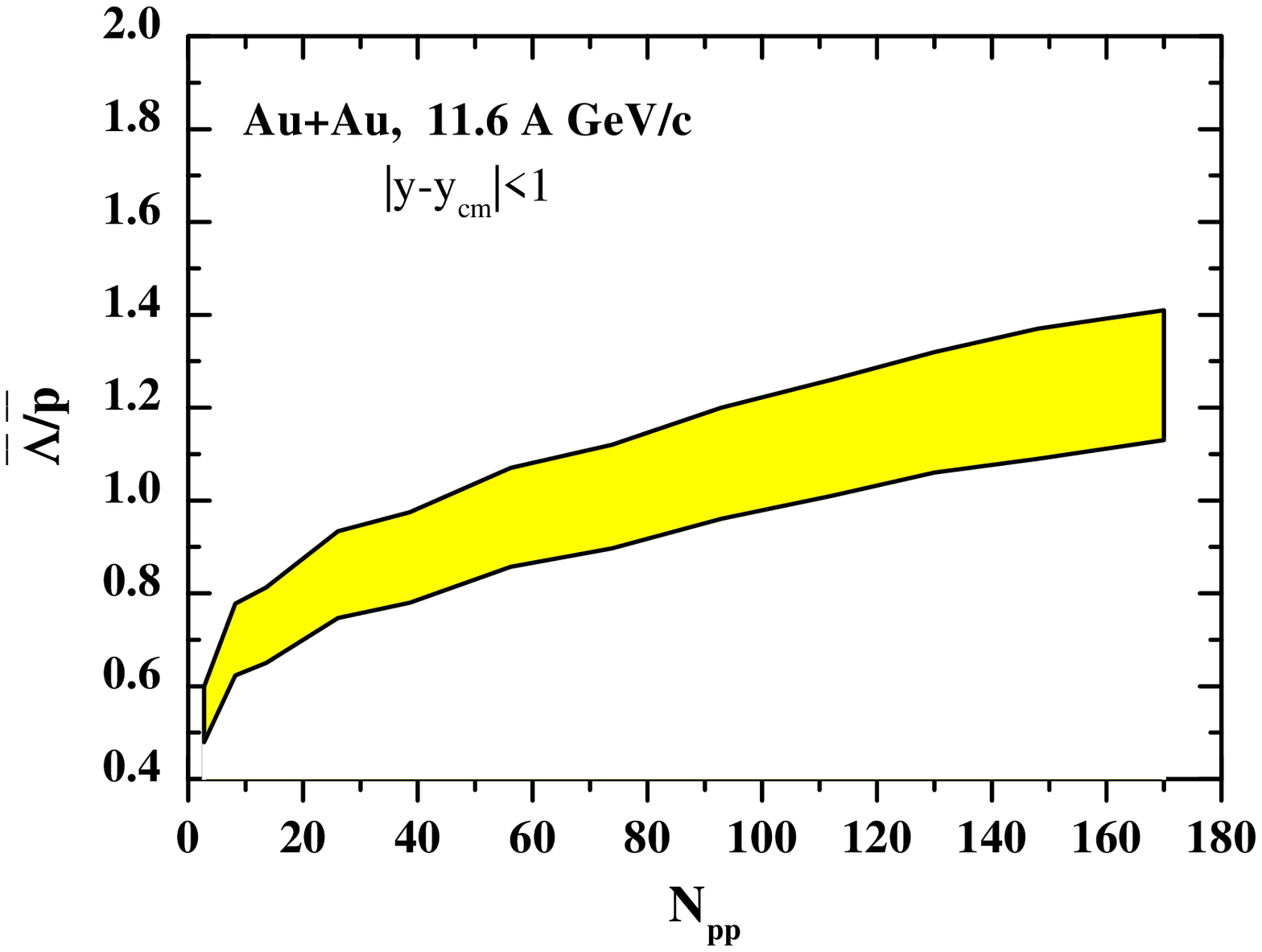,width=15cm}} \vspace*{-5cm}
\caption{The $\bar{\Lambda}/\bar{p}$ ratio as a function of
centrality for $Au+Au$ at 11.6 A$\cdot$GeV/c in the HSD approach
 within the approximation
(\ref{antihyp}). The shaded area corresponds to the uncertainty in
the statistics of the transport calculations for the different
particle abundancies at midrapidity.} \label{bild11}
\end{figure}


\begin{thebibliography}{99}
\bibitem{chamberlain} O. Chamberlain et al., Nuovo Cimento { 3} (1956) 447.
\bibitem{elioff} T. Elioff et al., Phys. Rev. { 128} (1962) 869.
\bibitem{dorfan} D. Dorfan et al., Phys. Rev. Lett. { 14} (1965) 995.
\bibitem{JINR} A. A. Baldin et al., JETP Lett. { 47} (1988) 137.
\bibitem{BEVALAC1} J. B. Carroll et al., Phys. Rev. Lett. { 62} (1989) 1829.
\bibitem{BEVALAC2} A. Shor, V. Perez-Mendez, K. Ganezer,
    Phys. Rev. Lett. { 63} (1989) 2192.
\bibitem{KEK} J. Chiba et al., Nucl. Phys. { A 553} (1993) 771c.
\bibitem{GSI} A. Schr\"oter et al., Nucl. Phys. { A 553} (1993) 775c.
\bibitem{Batko91} G. Batko, W. Cassing, U. Mosel, K. Niita,
    Phys. Lett. { B 256} (1991) 331.
\bibitem{Cass92} W. Cassing, A. Lang, S. Teis, K. Weber,
Nucl. Phys. { A 545} (1992) 123c.
\bibitem{Faess1}  S. W. Huang, G. Q. Li, T. Maruyama, A. Faessler,
    Nucl. Phys. { A 547} (1992) 653.
\bibitem{Danielewicz90} P. Danielewicz, Phys. Rev. { C 42} (1990)
1564.
\bibitem{Weise} E. Hernandez, E. Oset, W. Weise,
    Z. Phys. A 351 (1995) 99.
\bibitem{Schaffner91} J. Schaffner, I. N. Mishustin, L. M. Satarov,
    H. St\"ocker, W. Greiner, Z. Phys. { A 341} (1991) 47.
\bibitem{Koch} V. Koch, G. E. Brown, C. M. Ko, Phys. Lett. B 265 (1991) 29.
\bibitem{Teis94} S. Teis, W. Cassing, T. Maruyama, U. Mosel,
    Phys. Rev. C 50 (1994) 388.
\bibitem{Ko93} G. Q. Li, C. M. Ko, X. S. Fang, Y. M. Zheng,
    Phys. Rev. C 49 (1994) 1139.
\bibitem{RQMD} C. Spieles, A. Jahns, H. Sorge, H. St\"ocker,
    W. Greiner, Mod. Phys. Lett. A 27 (1993) 2547.
\bibitem{sibirtsev} A. Sibirtsev, W. Cassing, G. I. Lykasov,
    M. V. Rzjanin, Nucl. Phys. A 632 (1998) 131.
\bibitem{Cass99} W. Cassing, E. L. Bratkovskaya,
    Phys. Rep. 308 (1999) 65.
\bibitem{Ko96}  C. M. Ko, G. Q. Li, J. Phys. G 22 (1996) 1673.
\bibitem{ko1} C. M. Ko, X. Ge, Phys. Lett. { B 205} (1988) 195.
\bibitem{ko2} C. M. Ko, L. H. Xia, Phys. Rev. { C 40} (1989) R1118.
\bibitem{Wittmann} R. Wittmann, PhD thesis, Univ. of Regensburg,
1995; R. Wittmann, U. Heinz, {\it hep-ph}/9509328.
\bibitem{Dover} C. B. Dover, T. Gutsche, M. Maruyama, A. Faessler,
    Prog. Part. Nucl. Phys. 29 (1992) 87.
\bibitem{AGSall} L. Ahle et al., Nucl. Phys. A 610 (1996) 139c.
\bibitem{E877}  J. Barrette et al., Phys. Lett. B 485 (2000) 319.
\bibitem{AGSnew} B. B. Back et al., {\it nucl-ex}/0101008.
\bibitem{NA49} G. I. Veres and the NA49 Collaboration,
    Nucl. Phys. A 661 (1999) 383c.
\bibitem{Na49b} J. B\"achler et al., Nucl. Phys. A 661 (1999) 45c.
\bibitem{NAxx} F. Antinori et al., Nucl. Phys. A 661 (1999) 130c;
    R. Caliandro et al., J. Phys. G 25 (1999) 171.
\bibitem{NA57} F. Antinori et al., Nucl. Phys. A 681 (2001) 165c;
{\it hep-ex/0105049}.
\bibitem{Andersen} E. Andersen et al., Phys. Lett. B 449 (1999)
401.
\bibitem{AGS1} A. Jahns, H. St\"ocker, W. Greiner, H. Sorge,
    Phys. Rev. Lett. 68 (1992) 2895.
\bibitem{AGS2} S. H. Kahana, Y. Pang, T. Schlagel, C. B. Dover,
    Phys. Rev. C 47 (1993) 1356.
\bibitem{Kahana} Y. Pang, D. E. Kahana, S. H. Kahana, H. Crawford,
    Phys. Rev. Lett. 78 (1997) 3418.
\bibitem{Bleicher} M. Bleicher et al., Phys. Lett. B 485 (2000) 133.
\bibitem{WA97c} F. Antinori et al., Eur. Phys. J. C 11 (1999) 79.
\bibitem{Sorge} H. Sorge, Z. Phys. C 67 (1995) 479;
    Phys. Rev. C 52 (1995) 3291.
\bibitem{Werner} K. Werner, J. Aichelin, Phys. Lett. B 308 (1993) 372.
\bibitem{Carlos} M. A. Braun, C. Pajares, Nucl. Phys. B 390 (1993) 542;
    N. Armesto, M. A. Braun, E. G. Ferreiro, C. Pajares,
    Phys. Lett. B 344 (1995) 301; M. A. Braun, C. Pajares, J. Ranft,
    Int. Jour. Mod. Phys. A 14 (1999) 2689.
\bibitem{Carlos2} N. S. Amelin, N. Armesto, C. Pajares, D. Sousa,
    {\it hep-ph}/0103060.
\bibitem{Rafelski} P. Koch, B. M\"uller, J. Rafelski,
    Phys. Rep. 142 (1986) 167.
\bibitem{BM} P. Braun-Munzinger, I. Heppe, J. Stachel,
    Phys. Lett. B 465 (1999) 15.
\bibitem{becca} F. Becattini et al., Phys. Rev. C 64 (2001) 024901.
\bibitem{Redlich00} F. Becattini, J. Cleymans, A. Keranen, E. Suhonen,
    K. Redlich, {\it hep-ph}/0011322.
\bibitem{Red01} K. Redlich, {\it hep-ph}/0105104.
\bibitem{Redlich2} S. Hamieh, K. Redlich, A. Pounsi,
    Phys. Lett. B 486 (2000) 61.
\bibitem{Capella} A. Capella, Nucl. Phys. A 661 (1999) 502c.
\bibitem{Rapp} R. Rapp, E. Shuryak, Phys. Rev. Lett. 86 (2001) 2980.
\bibitem{Carsten} C. Greiner, S. Leupold, nucl-th/0009036.
%
\bibitem{Brat} E. L. Bratkovskaya, W. Cassing, C. Greiner et al.,
    Nucl. Phys. A 675 (2000) 661.
\bibitem{Bravina} L. V. Bravina et al., J. Phys. G 25 (1999) 351;
    Phys. Rev. C 62 (2000) 064906.
\bibitem{Ehehalt} W. Ehehalt, W. Cassing, Nucl. Phys. A 602 (1996) 449.
\bibitem{KLW1} K. Weber et al.,  Nucl. Phys. { A 539} (1992) 713;
    Nucl. Phys. { A 552} (1993) 571; T. Maruyama et al., Nucl.
    Phys. A 573 (1994) 653.
\bibitem{Mal} W. Botermans, R. Malfliet, Phys. Rep. 198 (1990) 115.
\bibitem{Cassing90} W. Cassing, V. Metag, U. Mosel, K. Niita,
    Phys. Rep. { 188} (1990) 363.
\bibitem{Cassing90c} W. Cassing, U. Mosel, Prog. Part. Nucl. Phys. 25 (1990) 235.
\bibitem{Casju} W. Cassing, S. Juchem, Nucl. Phys. A 665 (2000) 377;
    Nucl. Phys. A 672 (2000) 417; Nucl. Phys. A 677 (2000) 445.
\bibitem{Leupold} S. Leupold, Nucl. Phys. A 672 (2000) 475.
\bibitem{Byckling} E. Byckling, K. Kajantie, {\em Particle Kinematics}
          (John Wiley and Sons, London, 1973).
\bibitem{Lang93} A. Lang, H. Babovsky, W. Cassing, U. Mosel, H.-G. Reusch,
    K. Weber , J. Comp. Phys. { 106} (1993) 391.
\bibitem{LB} H. Schopper (Editor), Landolt-B\"ornstein, New
Series, Vol. I/12, Springer-Verlag, 1988.
\bibitem{PDG} Particle Data Group, Eur. Phys. J. C 15 (2000) 1.
\bibitem{Geiss} J. Geiss, W. Cassing, C. Greiner,
    Nucl. Phys. A 644 (1998) 107.
\bibitem{Wolf90} Gy. Wolf et al., Nucl. Phys. A 517 (1990) 615;
    Nucl. Phys. A 552 (1993) 549.
\bibitem{Batko3} G. Batko, J. Randrup, T. Vetter,
    Nucl. Phys. A 536 (1992) 786; Nucl. Phys. A 546 (1992) 761.
\bibitem{kodama} T. Kodama, S. B. Duarte, K. C. Chung et al.,
    Phys. Rev. C 29 (1984) 2146.
\bibitem{Bertsch88} G. F. Bertsch, S. Das Gupta, Phys. Rep. 160 (1988) 189.
\bibitem{Bass98} S. Bass et al., Prog. Part. Nucl. Phys. 41 (1998)
225.
\bibitem{Larionov} A. B. Larionov, W. Cassing, S. Leupold, U.
Mosel, {\it nucl-th}/0103019, Nucl. Phys. A, in press.
\bibitem{NA44} I. G. Bearden et al., Nucl. Phys. A 610 (1996) 175c;
    M. Caneta et al., J. Phys. G 23 (1997) 1865.
\bibitem{Zimanyi1} J. Zimanyi, in: S. Bass et al.,
    Nucl. Phys. A 661 (1999) 205c.
\bibitem{Zimanyi2} J. Zimanyi, T. S. Biro et al., {\it hep-ph}/9904501.
\bibitem{Sahu00} P. K. Sahu, W. Cassing, U. Mosel, A. Ohnishi,
    Nucl. Phys. A 672 (2000) 376.
\bibitem{Cass00a} W. Cassing, E. L. Bratkovskaya, S. Juchem, Nucl.
    Phys. A 674 (2000) 249.
\bibitem{E864n} T. A. Armstrong et al., Phys. Rev. C 59 (1999)
 2699.
\end{thebibliography}
\end{document}